\documentclass[aps,pre,floats, twocolumn,showpacs,superscriptaddress,
  reprint]{revtex4}
\usepackage{graphicx,epsfig}
\usepackage{graphics,dcolumn,bm,epic, eepic,fleqn,float}
\usepackage{amssymb,amsmath,amsfonts,multirow,rotate,color}
\usepackage{soul}
\usepackage{wasysym}
\usepackage{subfigure}

\usepackage{algorithm}
\usepackage{algorithmic}

\definecolor{Myorange}{cmyk}{0,0.42,1,0}

\newcommand{\avg}[1]{\langle #1 \rangle}

\newcommand{\lay}[1]{^{[#1]}}

\begin{document}

\title{Measuring and modeling correlations in multiplex networks}

\author{Vincenzo Nicosia}
\email{v.nicosia@qmul.ac.uk}
\affiliation{School of Mathematical Sciences, Queen Mary University of London, 
London E1 4NS, United Kingdom}  

\author{Vito Latora}
\affiliation{School of Mathematical Sciences, Queen Mary University of London, 
London E1 4NS, United Kingdom}  

\begin{abstract}
The interactions among the elementary components of many complex systems 
can be qualitatively different. Such systems are
therefore naturally described in terms of multiplex or
multi-layer networks, i.e.~networks where each layer stands for a
different type of interaction between the same set of nodes. There is
today a growing interest in understanding when and why a description
in terms of a multiplex network is necessary and more informative than
a single-layer projection. Here, we contribute to this debate by
presenting a comprehensive study of correlations in multiplex
networks. Correlations in node properties, especially degree-degree
correlations, have been thoroughly studied in single-layer
networks. Here we extend this idea to investigate and characterize
correlations between the different layers of a multiplex
network. Such correlations are intrinsically multiplex, and we first
study them empirically by constructing and analyzing several multiplex
networks from the real-world. In particular, we introduce
various measures to characterize correlations in the activity of the
nodes and in their degree at the different layers, and between
activities and degrees. We show that real-world networks exhibit
indeed non-trivial multiplex correlations. For instance, we find 
cases where two layers of the same multiplex network are
positively correlated in terms of node degrees, while other two layers
are negatively correlated. We then focus on constructing synthetic
multiplex networks, proposing a series of models to reproduce the
correlations observed empirically and/or to assess their relevance.
\end{abstract}


\maketitle

\section{Introduction}

Since  its origins, the new science of complex networks has been
primarily driven by the need to characterize the properties of
real-world systems~\cite{Newman2003rev,Boccaletti2006}.  The
introduction of new ideas and concepts in the field has been very
often associated to the availability of new, more accurate, or larger
data sets, and to the discovery of new structural properties of
complex systems from the real
world~\cite{Watts1998,Barabasi1999,Amaral2000,Barrat2004,Pastor-Satorras2001,
  Vespignani2001,Latora2001,Newman2001c,Newman2002,Girvan2002}. This
is the reason why a lot of interest has been recently devoted to the
study of multiplex networks, i.e. networks in which the same set of
nodes can be connected by means of links of qualitatively different
type or nature.  

Several data sets of real-world systems that can be represented and
studied as multiplex networks have appeared in the recent
literature~\cite{Szell2010,Cardillo2013_Emergence,Cardillo2013_Airports,Battiston2014},
and we expect that many more will arrive in the next few years.  The
first papers on the subject have focused on the characterization of
the structure of multiplex
networks~\cite{DeDomenico2013_tensor,DeDomenico2013_Centrality,
  Menichetti2014,Battiston2014, Sola2013, Cozzo2013_Clustering,
  Estrada2013,Corominas2013,Radicchi2013,
  Bianconi2014_component,Bianconi2013,Halu2013,
  DeDomenico2015structural}, and on modeling the basic mechanisms of
their
growth~\cite{Nicosia2013,Kim2013,Nicosia2014,Klimek2013,Fotouhi2015}. In
parallel to this, some effort has been also devoted on investigating
various kinds of dynamical processes on multiplex topologies,
including
diffusion~\cite{Gomez2013,DeDomenico2013_RW,Sole2013_Spectral,
  Halu2013_Pagerank,Battiston2015_biased}, epidemic
spreading~\cite{Granell2013,Zhao2013,Saumell2012, Buono2013,
  Min2013_epidemic, Cozzo2013_epidemic},
cooperation~\cite{GomezGardenes2012, Min2013_coop, Jiang2013,
  Wang2013}, opinion
formation~\cite{Diakonova2014,Chmiel2015,Battiston2015_ising,
  Diakonova2015}, and
percolation~\cite{Cellai2013,Baxter2013,Min2013_coop,Bianconi2014_percolation,Radicchi2015}.

There is today a general agreement on the fact that multiplex networks
represent the ideal framework to study a large variety of complex
systems of different nature. And there are already some numerical and
analytical results showing that the dynamics of processes on multiplex
networks is far richer than in networks with a single layer. A
comprehensive review of the main advances in this new vibrant field of
research can be found in a few recent survey
papers~\cite{Kivela2014,Boccaletti2014,Lee2015}.

In this Article we focus on an issue that has revealed of great
importance in single-layer networks, but has not yet been investigated
thoroughly in multiplex networks, i.e. that of
correlations~\cite{Parshani2010,Lee2012,Min2013_coop,Nicosia2014}.  In
networks with a single layer it has been found that there are
correlations in the properties of connected nodes. Namely, the degree
of a node can be either positively or negatively correlated with the
degree of its first neighbors. In the first case, the hubs of the
networks are preferentially linked to each others, while in the second
case they are preferentially connected to low-degree
nodes~\cite{Pastor-Satorras2001,Newman2002}. 

In multiplex networks the very same concept of correlations is far
richer than in a network with a single layer. In fact, on one hand it
is still possible to explore the standard degree-degree correlations
at the level of each layer of the network, but on the other hand it is
more interesting to introduce a truly multiplex definition of
correlations, for instance by looking at how a certain property of a
node at a given layer is correlated to the same or other properties of
the same node at another layer. We present here a complete and
self-consistent study of correlations of node properties in multiplex
networks. In doing this, we follow the usual steps of the typical
approach to complex networks: {\em i)} we first explore empirically
correlations in real multiplex networks, {\em ii)} we introduce
various measures to characterize and quantify correlations in
multiplex networks, {\em iii)} we propose a series of models to
reproduce the correlations found in real multiplex systems, or to
assess their relevance.

We find that multiplexity introduces novel levels of complexity. In
particular, in real-world multiplex networks the patterns of presence
and involvement of the nodes at the different layers are characterized
by strong correlations. And this has to be taken into account when it
comes to model such systems.

The Article is organized as follows.
In Sect.~\ref{why} we focus on two small real-world networks and we
use them as examples to explain why a description in terms of
multiplex networks captures more information on a system than a
single-layer projection.
In the remaining sections we study the structure of five real-world
multiplex networks (additional information about the networks are
provided in Appendix), with the main attention to the concept of
correlations.  In particular, in Sect.~\ref{sec:activity} we focus on
the patterns of node activity and involvement at the various layers.
We say that a node is active at a given layer if it has at least one
link at that layer, and we introduce various quantities to
characterize the distribution and the correlations of node
activities. We also investigate the activity correlations between
pairs of layers.  We find that real-world multiplex networks are quite
sparse, with only a few nodes active in many layers, and are
characterized by strong correlations: interestingly, the activity of a
node in a particular layer is very often correlated with its activity
in some other layer.

In Sect.~\ref{sec:activity_models} we introduce the first null-models
to assess the significance of the observed patterns of node
activity. In Sect.~\ref{sec:cartography} we investigate correlations
between the activity and the degree of the nodes of a multiplex
network, while in Sect.~\ref{sec:degree-degree} we show how to
quantify inter-layer degree correlations (degree correlations between
layers). In particular, we focus our attention on measuring
correlation in the node degrees of a pair of layers, either by using
the Spearman's rank correlation coefficient of the two degree
sequences, or by plotting, as a function of $k$, the average degree
$\bar{q}(k)$ at one layer of nodes having degree $k$ at the other
layer.  We find that there exist significant correlations among the
degree of the same node at different layers, and such correlations can
be either positive, meaning that nodes tend to have similar roles
across layers, or negative, meaning that nodes with a large degree in
one layer tend to have small degrees in another layer.

Finally, in Sect.\ref{sec:model_degree-degree}, we propose two
algorithms based on simulated annealing which allow to construct
multiplex networks with tunable inter-layer degree-degree
correlations, and in Sect.~\ref{sec:conclusions} we report our
conclusions.
The details on the five multiplex networks constructed from data sets
of biological, technological and social complex systems, and analyzed
in the paper, are describe in the Appendix. The networks and the
software implementations of the algorithms described in this paper are
available at~\cite{url}.

\begin{figure*}
  \begin{center}
    \includegraphics[width=6in]{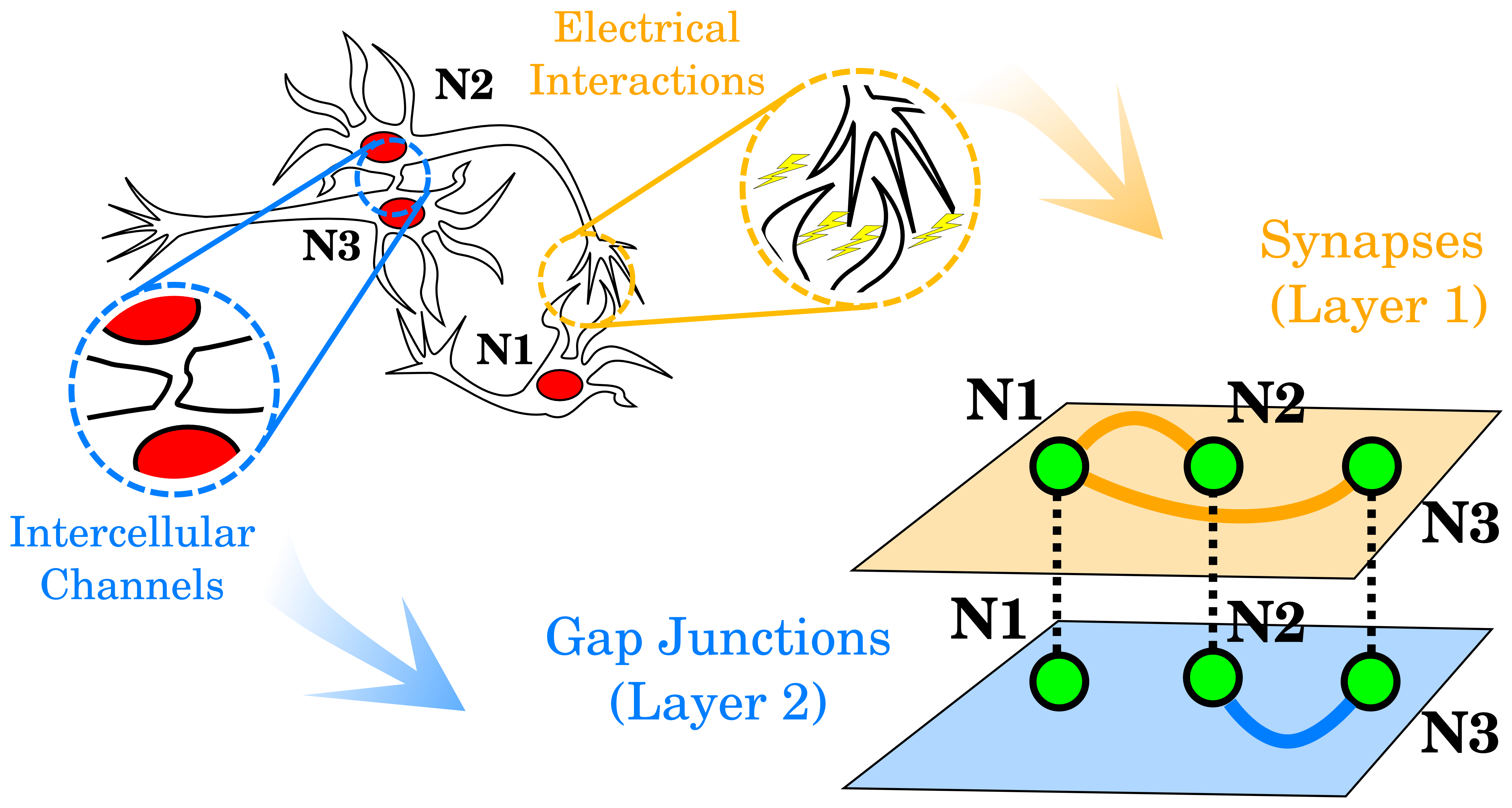}
  \end{center}
  \caption{(Color online) In a multiplex representation different types of
    relationships correspond to the distinct layers of a multi-layer
    network. For instance, in the case of the neural system of the C.elegans  
    two neurons can communicate either by means of electrical signals,
    which are propagated through synapses and neuronal dendrites, or
    by means of the diffusion of ions and small molecules, which
    travel through inter-cellular channels called gap junctions. The 
    two types of communication are encoded in the two
    layers of a multiplex network.}
  \label{fig:fig1}
\end{figure*}

\section{Why a description in terms of a multiplex network?}
\label{why}

The aim of this work is to identify, measure and model the different
kinds of correlations among node properties which can be found in a
multiplex network. For such a reason we constructed
\begin{table}
  \begin{center}
    \begin{tabular}{|l|rrr|}
      \hline
      Network & $N$ & $M$ & $\avg{N\lay{\alpha}}$\\\hline
      \hline
      C.elegans & 281 & 2 & 267 \\\hline
      BIOGRID & 54549 & 2 & 32143\\\hline
      Airlines - Africa & 235 & 84 & 9.8 \\\hline
      Airlines - Asia & 792 & 213 & 24.4\\\hline
      Airlines - Europe & 593 & 175 & 21.8 \\\hline
      Airlines - North America & 1020 & 143 & 24.9 \\\hline
      Airlines - Oceania & 261 & 37 & 14.1\\\hline
      Airlines - South America & 296 & 58 & 15.1\\\hline
      APS & 170385 & 10 & 43188 \\\hline
      IMDb & 2158300 & 28 &  229330\\\hline
      \hline
    \end{tabular}
  \end{center}
  \caption{Number of nodes $N$, number of layers $M$ and average
    number of active nodes $\avg{N\lay{\alpha}}$ of the multiplex
    networks analyzed in this study.}
\label{tableintro}
\end{table}
several multiplex networks from five data sets of real-world
systems. The systems we consider are: the nervous system of a
roundworm at the cellular level (C.elegans), a systems of interactions
between proteins (BIOGRID), the routes of continental airlines
(OpenFlight), the papers published in the journals of the American
Physical Society (APS), and the movies in the Internet Movie Database
(IMDb). These data sets are representative of the major classes of
complex systems, namely social, technological and biological, and
their sizes range from hundreds of nodes and just two kinds of
interactions in the case of C.elegans up to millions of nodes and
dozens of layers in IMDb. We provide in this way the reader with some
novel multiplex data sets, in addition to those already appeared in
the recent
literature~\cite{Cardillo2013_Emergence,Cardillo2013_Airports,Battiston2014,DeDomenico2015structural},
also showing that some well-known networks, such as the neural system
of the C.elegans, and the collaboration network of movie actors, can
indeed be better represented as multiplex networks. Basic
characteristics of the networks we have constructed, such as number of
nodes $N$, number of layers $M$ and average number of active nodes per
layer $\avg{N\lay{\alpha}}$ are shown in Table~\ref{tableintro}.
Additional details about the original data sets and the procedure used
to construct the networks can be found in Appendix. All the data sets
are available for download at~\cite{url}.

\begin{table}[!t]
  \begin{tabular}{|l|ccccc|}
    \hline
    Layer & $N\lay{\alpha}$ & $K\lay{\alpha}$ &
    $\avg{k\lay{\alpha}}$ & $N_C\lay{\alpha}$ & $S_1\lay{\alpha}, S_2\lay{\alpha}, S_3\lay{\alpha}$\\ \hline\hline
    \multicolumn{6}{|c|}{C.elegans}\\\hline
    Synapses        & 281 & 1962 & 13.9 & 2 & 279,2,-- \\ 
    Gap junctions   & 281 & 517 & 3.7 & 31 & 248,3,2 \\
    Aggregated & 281 & 2291 & 16.3 & 2 & 279,2,-- \\
    \hline
    \multicolumn{6}{|c|}{BIOGRID}\\\hline
    Genetic        & $12590$ & $203328$ & $32.3$ & $163$ & 9784, 1110, 979\\
    Physical       & 51697 & 299722 & 11.56 & 664 & 50213, 20, 20 \\
    Aggregated & 54549 & 500239 & 18.34 & 607 & 52879, 304, 20 \\
    \hline
  \end{tabular}
  \caption{The number of active nodes $N\lay{\alpha}$, the number of
    edges $K\lay{\alpha}$, the average degree $\avg{k\lay{\alpha}}$,
    the number of components $N_C\lay{\alpha}$ and the size of the
    three largest components at the two layers of the C.elegans neural
    network, and at the two layers of the BIOGRID protein interaction
    network. We report for reference also the values corresponding to
    the networks obtained by aggregating the two layers together.}
  \label{tab:celegans_biogrid}
\end{table}

Before moving to the main topic of our work and to the various ways of 
formalizing and measuring correlations in a multiplex network, we focus 
in this section on what we gain by studying a system as a
multiplex network, instead of aggregating together its different
layers.  We will do this by considering two of the real-world
multiplex networks we have introduced, namely the two two-layer
biological systems reported in Table.~\ref{tab:celegans_biogrid}: the
C.elegans neural system and the BIOGRID protein-gene interaction
network.  The first thing we notice from a component analysis of such
systems at the two layers is that not all the nodes are connected in
both layers. For instance, the synaptic layer (Syn) of the C.elegans
neural network consists of two connected components of $279$ and $2$
nodes, while in the gap-junction layer (Gap) we observe a large
connected component containing 248 nodes, two small components
respectively with three and two nodes, and 28 isolated nodes.
Secondly, the two layers of the C.elegans have largely different densities. The
synaptic layer has an average degree equal to $\avg{k\lay{Syn}}=13.9$, 
while the gap-junction layer has $\avg{k\lay{Gap}}=3.7$ only. Additionally,
each node can play a very different role in the two layers. 
As an example, we report in
Table~\ref{tab:celegans_rank} the list of the top ten nodes ranked by
degree centrality in each of the two layers. Despite some nodes have similar
positions in the two rankings (e.g., AVAL, AVAR, AVBR), in general a
node with a high degree in the synaptic layer might have just a few
links in the other layer, as in the case of node AVDR, which is ranked
fourth in the synaptic layer, with 53 edges, but has only 4 edges in
the gap-junction layer. For reference, we also report in the same
table the ranking induced by the degree on the aggregated graph, which
is in turn different from the rankings corresponding to the two
single layers, especially from that at the gap-junctions level. 
\begin{table}
  \begin{tabular}{|c|cc|cc|cc|}
    \hline
    rank & Syn & $k\lay{Syn}$ & Gap & $k\lay{Gap}$ & Syn+Gap & $k$\\
    \hline
    1 & AVAR & 85 & AVAL & 40 & AVAL & 123\\
    2 & AVAL & 83 & AVAR & 34 & AVAR & 119 \\
    3 & AVBL & 56 & AVBR & 29 & AVBR & 80 \\
    4 & AVDR & 53 & AVBL & 24 & AVBL & 80 \\
    5 & PVCL & 52 & RIBR & 17 & PVCR & 60 \\
    6 & AVBR & 51 & RIBL & 17 & PVCL & 60 \\
    7 & AVER & 50 & AVKL & 14 & AVDR & 57\\
    8 & AVEL & 50 & RIGL & 14 & AVER & 56\\
    9 & PVCR & 49 & VA08 & 11 & AVEL & 55\\
    10 & DVA & 48 & RIGR & 11 & DVA  & 53\\
    \hline
  \end{tabular}
  \caption{The nodes ranked in the first ten positions according to
    their degree at the synapse layer, at the gap-junction layer and
    at the single-layer network obtained by aggregating the two
    layers.  Notice that some neurons are present in one of the two
    layer-based ranking and not in the other, e.g. PVCL and RIBR,
    indicating node can play different roles at the two
    layers. Moreover, also the ranking based on the degree of the
    aggregated network is different from the rankings at the two
    layers.}
  \label{tab:celegans_rank}
\end{table}

\begin{figure}
  \begin{center}
    \includegraphics[width=3in]{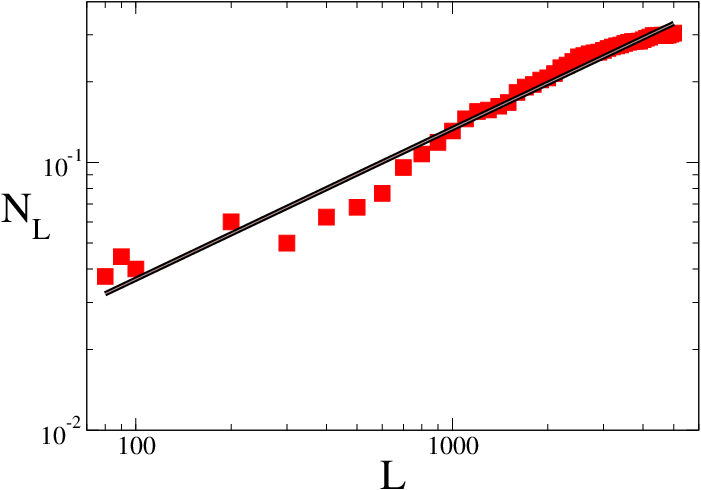}
  \end{center}
  \caption{(Color online) The fraction $N_L$ of nodes which appear in
    the top $L$ positions according to degree in both layers (Phys and
    Gen) of the BIOGRID network (squares) scales approximately as a
    power law $N_L \sim L^{0.56}$ (solid line, $r^2=0.96$). In
    particular, less than $20$ nodes appear in both rankings up to
    $L\simeq 300$, meaning that there is almost no correlation between
    the degrees of the a node at the two layers, and that it is very
    unlikely that a node is a hub on both Gen and Phys.}
  \label{fig:num_correspondence_BIO}
\end{figure}

Also the two layers of the BIOGRID network, respectively representing
physical (Phys) and genetic (Gen) interactions among proteins, have
radically different structures. First of all, the two layers have a
different number of non-isolated nodes and a different distribution of 
the sizes of connected components, with Phys having $N=51697$ non-isolated
nodes, while Gen only $N=12590$. Of these nodes, only $9738$ are
non-isolated on both layers, meaning that more than $80\%$ of the
nodes are active in just one of the two layers, and not in the
other. Despite having a smaller number of non-isolated nodes, the Gen 
layer is much denser than Phys, with an average degree
$\avg{k\lay{Gen}}\simeq 32$ compared to $\avg{k\lay{Phys}}\simeq
11$. Also in this case there is no correspondence between the hubs at the
two layers, as shown by the plot in 
Fig.~\ref{fig:num_correspondence_BIO} which reports the fraction $N_L$
of those nodes which are found in the top-L ranking of both layers
according to the degree. Notice that $N_L$ is much smaller than $10\%$
for a wide range of values of $L$ (i.e., up to $L\simeq 600$), meaning
that if a node is a hub on one layer there is a quite small
probability that it will also be a hub on the other layer. This result
is due to the fundamental difference between physical
interactions, which produce new protein compounds, with respect to genetic
interactions, which trigger the production of other proteins.

Summing up, if we take into account the multi-layered nature of the
C.elegans neuronal network and of the BIOGRID protein interaction network
we discover new structural patterns, and in particular a poor
correspondence of the roles of a node across layers, with a large
fraction of the nodes being isolated at least in one of the two
layers. These results suggest that representing a system 
as a multiplex network allows to retain important information, 
since multi-layer real-world systems are often 
characterized by non-trivial patterns of node involvement
across layers. In the rest of the paper we propose some metrics to
quantify these patterns, and we introduce a few models to reproduce 
and to assess their significance.

\begin{figure*}[!t]
  \begin{center}
    \includegraphics[height=2in]{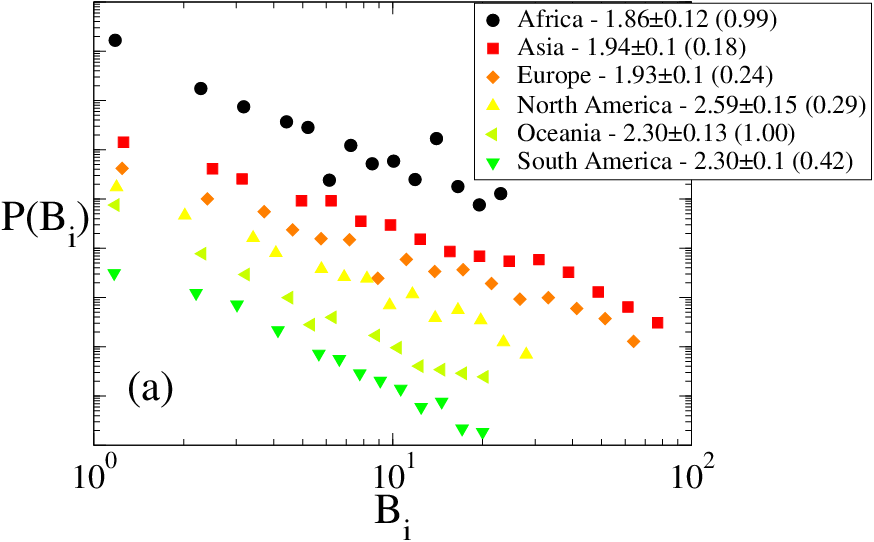}
    \includegraphics[height=2in]{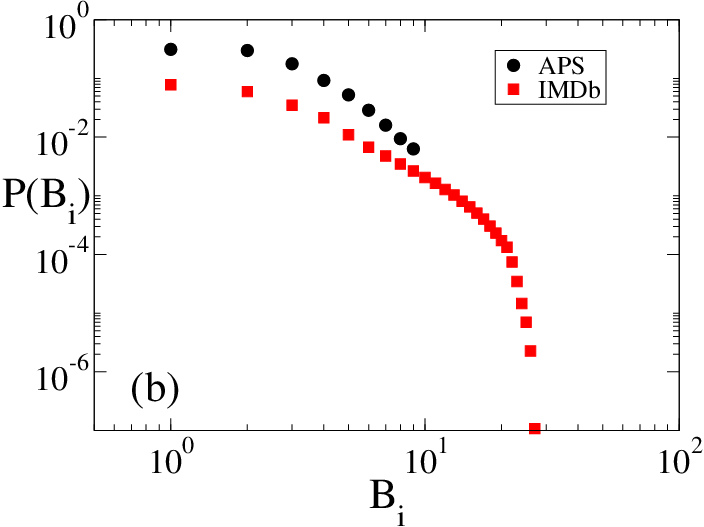}
  \end{center}
  \caption{(color online) Distributions of node-activity for (a) the
    six multiplex networks of continental airlines and for (b) APS and
    IMDb. In all airline networks $P(B_i)$ can be fitted by power-laws
    with exponents ranging from $1.8$ to $2.3$ (the exponents,
    together with the corresponding p-values in parenthesis, are
    reported in the legend).  This means that the typical number of
    layers in which a node is active is subject to unbounded
    fluctuations. The plots in panel (a) have been vertically
    displaced to enhance readability.}
  \label{fig:multi_activity_real}
\end{figure*}

\section{Correlations of node activity}
\label{sec:activity}

Let us consider a multiplex network with $N$ nodes and $M$ layers.
Such a network can be naturally described by giving a set of $M$
adjacency matrices, one for each layer, $\{A\lay{1}, A\lay{2}, \ldots,
A\lay{M}\}\in \mathbb{R}^{N\times N \times M}$, so that the element
$a\lay{\alpha}_{ij}=1$ if node $i$ and $j$ are connected at layer
$\alpha$, while $a\lay{\alpha}_{ij}=0$ otherwise. Notice that in this
particular kind of multiplex networks, a node $i$ effectively consists
of $M$ replicas, one for each layer, and inter-layer connections among
these replicas have no explicit cost
associated~\cite{Battiston2014}. These multiplexes are sometimes
referred in the literature as \textit{colored-edge
  networks~\cite{Kivela2014}.} In this framework, the properties of
the nodes are represented by vectorial variables. For instance, we can
associate to each node $i$ of the multiplex a {\em multi-degree},
i.e.~a $M$-dimensional vector
\begin{equation}
  \bm{k}_i = \left\{k\lay{1}_{i}, k\lay{2}_i, \ldots, k\lay{M}_i\right\}
\end{equation}
such that $k\lay{\alpha}_i$ denotes the degree of $i$ at layer
$\alpha$. A node can in fact participate with a different number of
edges to each layer, and can also be isolated in some of the
layers. Intuitively, the presence and number of edges incident in a
node is a first indication of the activity or importance of that node
at that layer.  But there is another level of complexity, typical of
multiplex structures, which is related to the importance or role of
one layer with respect to another in terms of the fraction of
connected nodes and of the relative number of edges of a certain
kind. For example, in the APS multiplex the number of active nodes in
the two Condensed Matter layers (layer 6 and layer 7) account for more
than one third of the total number of active nodes at all layers,
while the number of edges connecting authors working in General
Physics, Particle Physics, Nuclear Physics and Astronomy account for
more than 99\% of all the edges in the multiplex (see
Table~\ref{tab:APS} for details).
The additional complexity added by the presence of multiple layers
allows for the exploration of several kinds of structural
properties. In particular, we are interested here in detecting,
quantifying and modeling the existence of correlations of node
activity across layers (vertical analysis) and of correlations among
layer structures (horizontal analysis). To this aim, we define in the
following some basic quantities which characterize, respectively, the
activity of nodes and layers.

\subsection{Node activity}

We say that node $i$, with $i=1,2,\ldots,N$,  
is \textit{active} at layer $\alpha$ if
$k\lay{\alpha}_i>0$. We can then associate to each node $i$ a 
{\em node-activity vector} 
\begin{equation}
  \bm{b}_i = \left\{b\lay{1}_i, b\lay{2}_i, \ldots, b\lay{M}_i\right\}
\end{equation}
where
\begin{equation*}
b\lay{\alpha}_i=1 - \delta_{0, k\lay{\alpha}_i}
\end{equation*}
i.e., $b\lay{\alpha}_i=1$ if node $i$ has at least one edge at layer
$\alpha$, and is 0 otherwise. We call \textit{node-activity} $B_i$ of
node $i$ the number of layers on which node $i$ is active:
\begin{equation}
  B_i = \sum_{\alpha}b\lay{\alpha}_i
\end{equation}
By definition we have $0 \le B_i \le M$.  Notice that the
node-activity vector $\bm{b}_i$ provides a compact, yet incomplete
(because it does not take into account the number of links)
representation of the involvement of node $i$ at the different layers
of the multiplex.  However, we will show that it contains useful
information.

\textit{Distribution of node-activity. --- } In
Fig.~\ref{fig:multi_activity_real} we report the distributions of
node-activity obtained for the multiplex networks constructed from
OpenFlight, APS and IMDb. Interestingly, in the airport networks the
distributions are well fitted by a power-law function $P(B_i)\sim
B_i^{-\delta}$, with exponents $\delta$ in the range $[1.8, 2.4]$. The
values of the exponents were obtained through the maximum-likelihood
estimator~\cite{Clauset2007}. The most heterogenous distribution is
that of the African airplane multiplex network ($\delta\simeq 1.86$),
reported as black circles in Fig.~\ref{fig:multi_activity_real}(a),
while the two most homogeneous distributions are those of the airline
networks of South America and Oceania (both characterised by
$\delta\simeq 2.3$).
The power-law behaviour of node-activity indicates that there is no
meaningful typical number of layers on which a node is active, since
for $\delta < 3.0$ the fluctuations on this number are unbound as $M$
grows.  A scale-free distribution of node-activity in the airport
multiplex networks indicates that the majority of airports usually
tend to be connected only by a relatively small number of airlines
(between $68\%$ and $89\%$ of all the airports in each multiplex are
active in less than $5$ layers), but some ``outliers'' exist which are
connected by a relatively large number of different airlines (at least
one airport in each multiplex is active in $10\%$ to $30\%$ of the
layers).  Similar considerations can be made for APS and IMDb, where
the vast majority of authors and actors are active in just one or a
few layers, while a few outliers are found active in almost all
layers.

In the same spirit of what is done in single-layer networks, where
nodes having a relatively high number of connections in a network are
called \textit{hubs}, we call \textit{multi-active hubs} those outlier
nodes of a multiplex which are active in a large fraction of
layers. However, as we will better see in
Section~\ref{sec:cartography}, in real-world systems node-activity is
not strictly correlated to the total number of edges incident in a
node, so that a node might be a multi-active hub without being a hub
in the classical sense (of having many links) in any of the layers. In
particular, there exist nodes having, at the same time, a large number
of incident edges and a small node-activity (e.g., they might be
active in just a few layers, or even in one layer only), and also
nodes having a relatively small number of edges which are instead
active on almost all layers.

\textit{Distribution of node-activity vectors. --- } The
node-activity $B_i$ accounts only for the number of layers
at which node $i$ is active, discarding any information about which are
these layers. As a matter of fact, two nodes $i$ and $j$ might have the
same value of node-activity but they can be involved in different
layers. So it is interesting to look also at how the 
node-activity vectors ${\bf b}_i, i=1,2,\ldots,N$,  
are distributed, to see the relevant frequency of 
different node-activity patterns.
First of all it is important to notice that the actual number of
distinct node-activity vectors observed in a multiplex can in general
be much smaller than the total possible number of such vectors, which
is equal to $2^M-1$ (if we take into account only nodes that are
active on at least one layer). For instance, while in the APS
multiplex we observe $981$ out of the $1023$ possible node-activity
vectors (with an average of $173.6$ nodes having the same vector), in
IMDb we observe only around $123000$ out of more than $2.6\times
10^{8}$ possible vectors (with an average of around $17.4$ nodes
having the same vector).

\begin{figure}
  \begin{center}
    \includegraphics[width=3in]{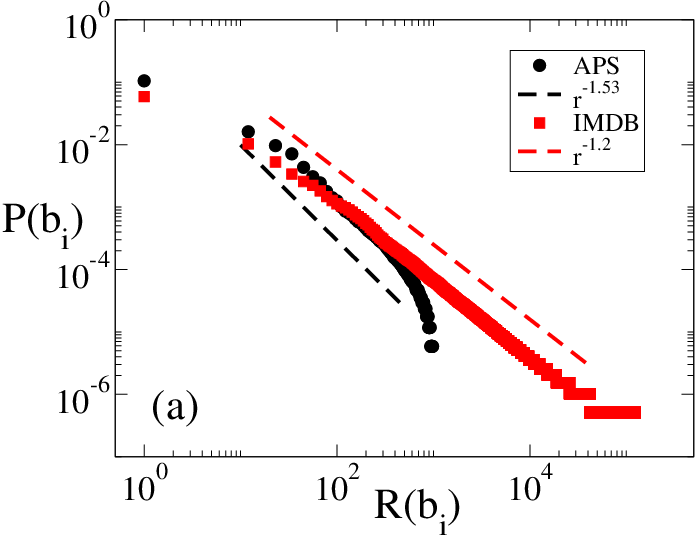}\\
    \includegraphics[width=3in]{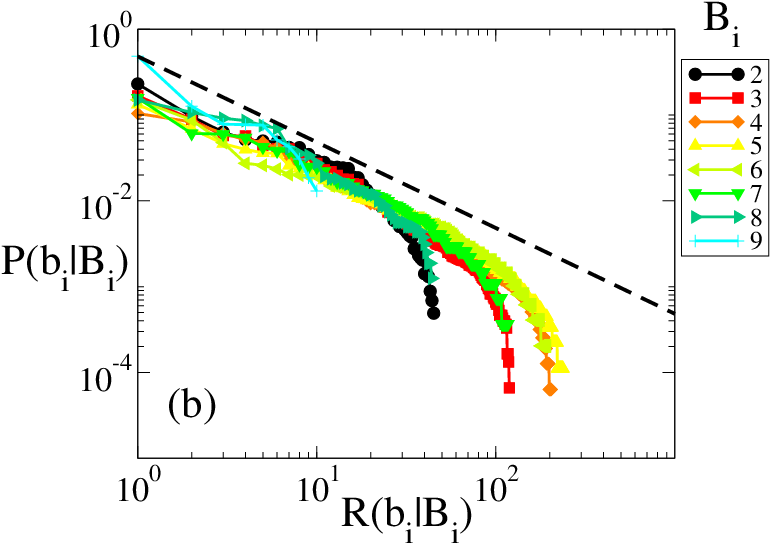}\\
    \includegraphics[width=3in]{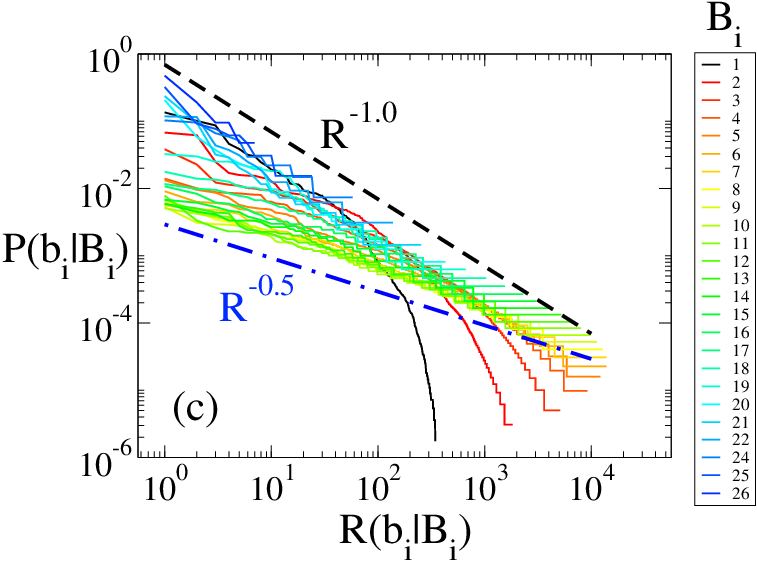}
  \end{center}
  \caption{(color online) (a) The Zipf's plot of the node-activity
    vectors is a power-law, both for APS and for IMDb. Also the rank
    distribution $P(\bm{b}_i|B_i)$ restricted to nodes having a given
    value of node-activity $B_i$, for (b) APS and (c) IMDb, are
    power-laws with exponential cut-off. The esponents of the
    power-laws range between $0.5$ (dot-dashed blue line) and $1.0$
    (dashed black line).}
  \label{fig:bi_rank_real}
\end{figure}

In Fig.~\ref{fig:bi_rank_real}(a) we show the Zipf's plot of the
node-activity vectors for the APS and IMDb multiplex networks. In both
cases the distribution of $\bm{b}_i$ is a power-law (with a clear
exponential cut-off in the case of APS), with an exponent respectively
equal to $1.53$ and $1.2$. This means that the majority of the nodes
have similar activity patterns, with the highest values of $P({\bf
  b}_i)$ always corresponding to nodes active on just one or two
layers, while some other node-activity vectors are more rare.
This result is also confirmed by Fig.~\ref{fig:bi_rank_real}(b) and
\ref{fig:bi_rank_real}(c) where we report, respectively for APS and
IMDb, the rank distributions, i.e., the Zipf's plots of the
probability $P(\bm{b}_i|B_i)$ of node-activity vectors $\bm{b}_i$
restricted to nodes active on exactly $B_i$ layers, as a function of
the rank $R(\bm{b}_i|B_i)$.

The various curves correspond to different values of $B_i$. Notice
that in general $P(\bm{b}_i|B_i)$ is heterogeneous and is a power-law
for the large majority of values of $B_i$. This means that a large
fraction of the nodes having the same value of node-activity share
also the same activity pattern across layers, while some outlier nodes
have quite peculiar activity patterns.  In the case of IMDb, for
instance, of all the actors who have worked on exactly two genres,
around $20\%$ are specialised in Short and Drama (layers 23 and 9) or
Short and Comedy (layer 23 and 6), while only one actor has acted both
on Fantasy and War movies (respectively layer 11 and 27) and only two
have acted both in an Adult movie and in a Family movie.
\begin{figure}[!ht]
  \begin{center}
    \includegraphics[width=3in]{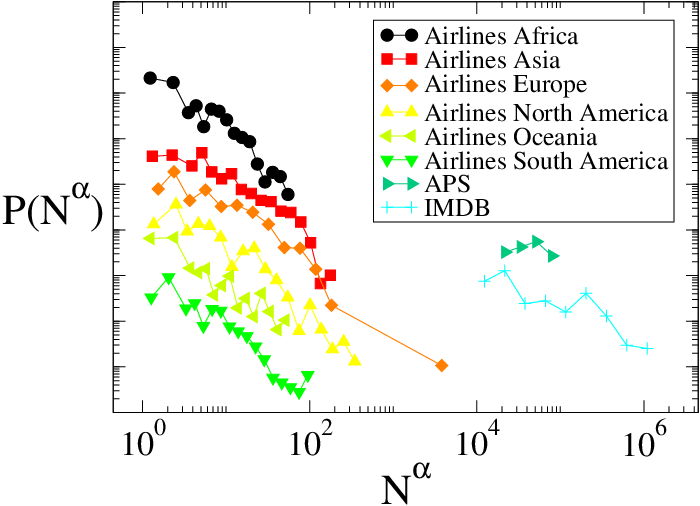}
  \end{center}
  \caption{(color online) Distribution of layer-activity for the
    continental airline networks, APS and IMDb. In the six multiplex
    of continental airlines, which consist of $O(10^2)$ layers,
    $P(N^{\alpha})$ has a clear power-law shape. A somehow
    heterogeneous behaviour is also observed for IMDb, although the
    number of layers is not large enough to allow a meaningful
    fit. The plots were vertically displaced to enhance readability.}
  \label{fig:layer_activity_real}
\end{figure}

\subsection{Layer activity}

The activity of a given layer $\alpha$, with $\alpha=1,2,\ldots,M$,
depends on the patterns of node activities at that layer, and can be
represented by the \textit{layer-activity vector}:
\begin{equation}
  {\bf d} \lay{\alpha} = \left\{b\lay{\alpha}_{1}, b\lay{\alpha}_{2}, \ldots,
  b\lay{\alpha}_{N}\right\}.
\end{equation}
We define the \textit{layer-activity} of layer $\alpha$ as the number
$N\lay{\alpha}$ of active nodes in $\alpha$, which is equal to the
number of non-zero elements of ${\bf d} \lay{\alpha}$:
\begin{equation}
  N\lay{\alpha} = \sum_i b\lay{\alpha}_i
\end{equation}
By definition we have $0 \le  N\lay{\alpha} \le N$. 

\textit{Distribution of layer-activity. --- } In
Fig.~\ref{fig:layer_activity_real} we show the distributions of
$N\lay{\alpha}$ for all the multiplex networks with more than two
layers. Interestingly, as found for $B_i$, also the distribution of
$N\lay{\alpha}$ is heterogeneous, and has a marked power-law behavior
for the continental airlines networks, which have a larger number of
layers. This confirms that, not only the activity of nodes across
layers is heterogeneous, but also that not all layers have the same
importance in the overall organisation of the system. For instance, a
large fraction of all the layers of the continental airlines
multiplexes have no more than $N\lay{\alpha}=10$ active
nodes. However, some layers contain up to a few hundred active nodes
(which account for $10\%$ up to $30\%$ of all the nodes). This means
that, on average, the removal of one layer at random from the system,
i.e., the removal of all the routes operated by one airline company,
will cause only minor disruptions, but in some specific cases such a
removal might break the system apart.

\begin{figure*}[!ht]
  \begin{center}
    \includegraphics[width=2.1in]{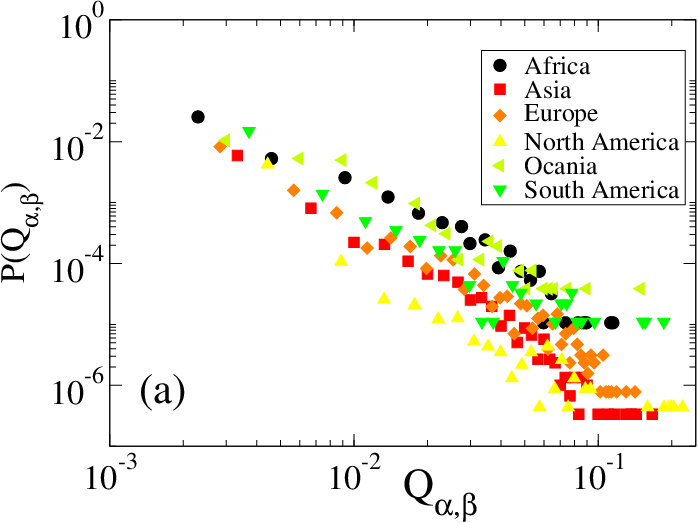}
    \includegraphics[width=1.95in]{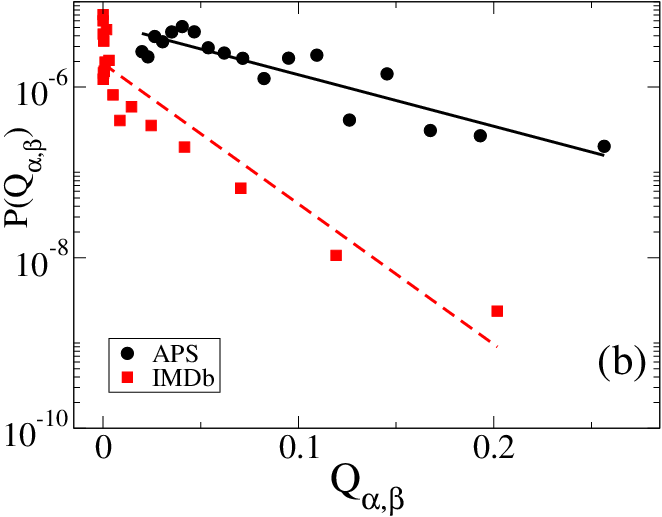}
    \includegraphics[width=2.6in]{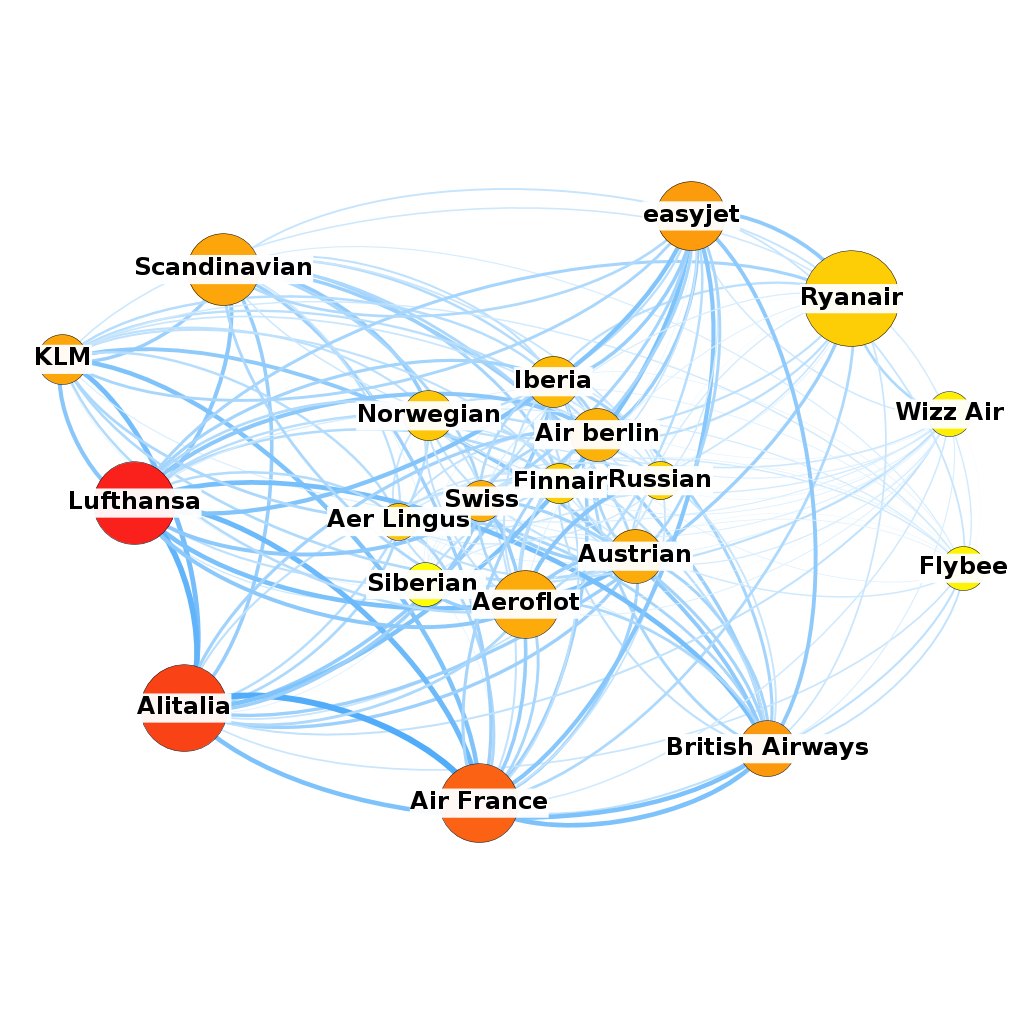}
  \end{center}
  \caption{(color online) The pairwise multiplexity has a power-law
    behavior in (a) airline networks, while it is exponential in (b)
    APS and IMDb. In panel (c) we report a graph of the first 20
    airlines in Europe by number of covered airports. Each node of
    this graph represents a layer of the original multiplex network,
    while the weight of the edge connecting two nodes is proportional
    to the fraction of nodes present in both layers. The size of a
    node is proportional to the number of airports in which the
    corresponding company operates, while the color (from yellow to
    red) corresponds to the node strength, which in this case is
    proportional to the total node overlap with other airlines.}
  \label{fig:multiplexicity_real}
\end{figure*}

\textit{Correlations of layer-activity. --- } We define here some
simple measures to detect and characterize the correlations among
layer activities. The first measure we propose is the \textit{pairwise
  multiplexity} $Q_{\alpha, \beta}$ of two layers $\alpha$ and $\beta$
defined as:
\begin{equation}
  Q_{\alpha, \beta} = \frac{1}{N} \sum_{i} b\lay{\alpha}_i
  b\lay{\beta}_i
\end{equation}
Notice that this quantity is equal to the fraction of nodes of the
multiplex which are active on both layers $\alpha$ and $\beta$, and
therefore takes values in the range $[0,1]$. The more similar the
activity pattern of the nodes at two layers, the higher the
multiplexity of two layers is.  The distribution of the values of the
pairwise multiplexity $P(Q_{\alpha,\beta})$ among all the possible
pairs of layers $\alpha$ and $\beta$ is reported in
Fig.~\ref{fig:multiplexicity_real}(a)-(b), respectively for the
continental airports networks and for APS and IMDb. We first notice
that in all the multiplex networks considered only a relatively small
fraction of nodes are active at the same time on at least two
layers. In particular, in the case of continental airlines the
multiplexity has a broad distribution, so that the majority of couples
of layers have less than $1\%$ of the nodes in common, while in a few
cases the multiplexity can be as high as $20\%$. Also in APS and IMDb
the values of pairwise multiplexity are usually below $20\%$, but in
this case the distributions exhibit an exponential decay, indicating
that there exists a typical scale of pairwise layer multiplexity.

The different behaviour of $P(Q_{\alpha,\beta})$ in the airport
networks, with respect to the collaboration networks is probably due
to the different meaning of activity at each layer, and also to the
different dynamics of node activation in the two cases. In particular,
for the airport network, we expect that the competition between
airlines operating in the same area produces a small overlap in the
activity pattern of the corresponding layers. This is clear shown in
Fig.~\ref{fig:multiplexicity_real}(c), where we report the graph
representing relationships among the first 20 airlines in Europe
operating in the largest number of airports. In this graph each node
represents a layer of the original multiplex network, its size is
proportional to the corresponding value $N\lay{\alpha}$, the width of
an edge is proportional to the pairwise multiplexity of the
corresponding layers, i.e. the fraction of nodes active on both
layers, and the color of nodes indicates the total multiplexity,
i.e. the sum of the values of pairwise multiplexity incident on a node
(red is maximum, yellow is minimum).  Notice that national companies,
like Lufthansa, Alitalia and Air France, tend to have a large overlap
with other airlines, i.e. to serve similar sets of airports, while
low-cost airlines, like easyJet, Ryanair, Wizz Air and Flybee,
systematically tend to avoid overlaps with other companies.
The relatively small values of pairwise multiplexity found in these
real-world multiplex networks may have an impact on the dynamics of
processes occurring over them, such as opinion formation, epidemic
spreading, percolation or
immunization~\cite{Radicchi2015,Diakonova2015}. Indeed, since only a
relatively small fraction of nodes are active on two layers at the
same time then the removal of just a few of these nodes might result
in a massive disruption of the multiplex network, and can thus slow
down dramatically either the spreading of an epidemic or the diffusion
of information. This aspect has to be properly taken into account when
considering dynamical processes on multiplex networks.

\begin{figure}[!t]
  \begin{center}
    \includegraphics[width=3in]{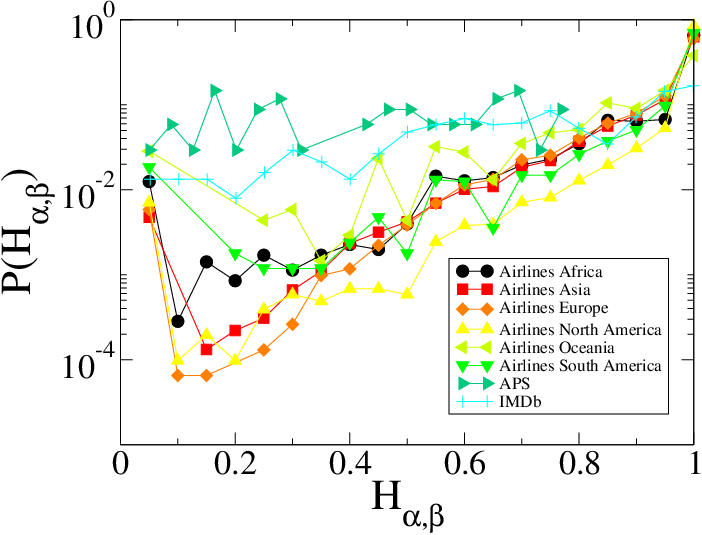}
  \end{center}
  \caption{(color online) The distribution of the normalised Hamming
    distance $H_{\alpha, \beta}$ between all the possible pairs of
    layers on various multiplex networks. Notice that $P(H)$ increases
    exponentially for the continental airlines networks.}
  \label{fig:hamming_real}
\end{figure}

Another measure to quantify the relative overlap between two layers at
the level of node activity is the normalised Hamming distance between
the two corresponding layer-activity vectors: 
\begin{equation}
  H_{\alpha,\beta} = \frac{\sum_{i}b\lay{\alpha}_i(1-b\lay{\beta}_i) +
    (1-b\lay{\alpha}_i)b\lay{\beta}_i}{\min(N\lay{\alpha} +
    N\lay{\beta}, N)}
\end{equation}
$H_{\alpha,\beta}$ is equal to the number of differences in the
activities of the two layers divided by the maximum possible number of
such differences, and takes values in $[0,1]$. In particular,
$H_{\alpha,\beta}=0$ if ${\bf d} \lay{\alpha}= {\bf d} \lay{\beta}$,
while $H_{\alpha,\beta}=1$ when all the active nodes at layer $\alpha$
are not active at layer $\beta$.
In Fig.~\ref{fig:hamming_real} we report the distributions of
$H_{\alpha,\beta}$ for the continental airlines, for APS and for
IMDb. In all the networks considered the measured values of
$H_{\alpha, \beta}$ are distributed throughout the whole $[0,1]$
range.  However, in the continental networks the distributions have an
increasing exponential behaviour, meaning that the normalised Hamming
distance is quite large for the vast majority of layer pairs, in
accordance with the observation that airports generally have small
node-activity (Fig.~\ref{fig:layer_activity_real}(a)). Conversely, for
APS and IMDb the distributions are more homogeneous. It is interesting
to notice that in all the systems around $1\%$ of the layer pairs have
a normalised distance smaller than $0.05$, corresponding to large
overlaps of node activity.
\begin{figure*}[!t]
  \begin{center}
    \includegraphics[width=3in]{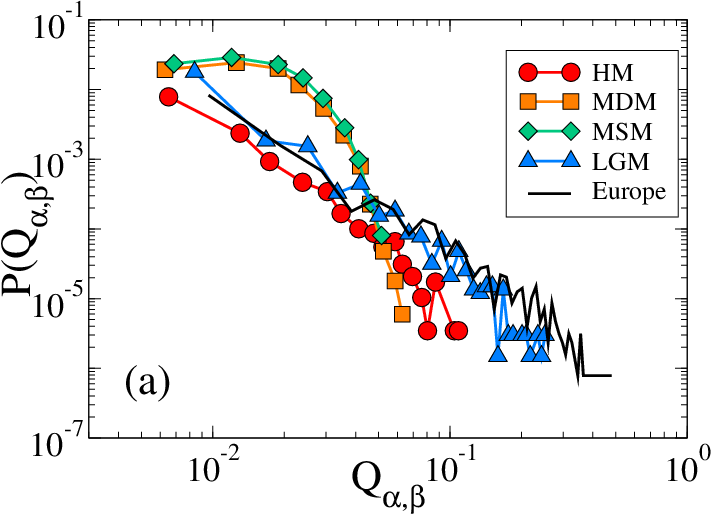}
    \includegraphics[width=3in]{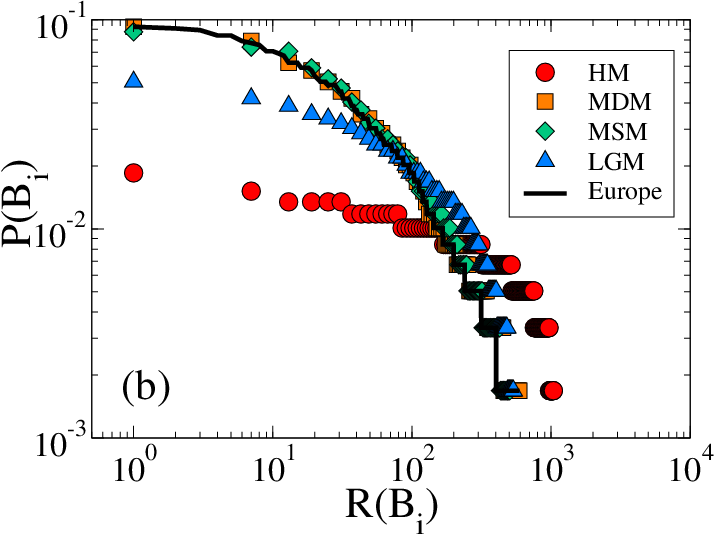}
  \end{center}
  \caption{(color online) Distribution of pairwise multiplexity (a)
    and Zipf's plot of node-activity (b) for the European airlines
    multiplex network (solid black line) and the corresponding
    synthetic networks obtained by four different models, namely: HM
    (red circles), MDM (orange squares), MSM (green diamonds) and LGM
    (blue triangles). Notice that LGM fits well the distribution of
    pairwise multiplexity, and performs better than HM in reproducing
    the rank distribution of node-activity. The shape of $P(B_i)$ in
    MDM and MSM is identical to that of the original multiplex by
    construction.}
  \label{fig:airports_real_models}
\end{figure*}

\section{Models of node and layer activity}
\label{sec:activity_models}

The empirical results of Section~\ref{sec:activity} suggest that the
patterns of node and layer activity in real-world multiplex networks
can be quite heterogeneous. In general, real-world multiplex systems
tend to be quite \textit{sparse}, meaning that the majority of nodes
participate to only a small subset of all the layers, and given two
layers only a small fraction of their nodes are active on both. It is
therefore natural to ask whether similar patterns might naturallly
arise from a random distribution of node activity across layers or
not.  Or, in other words, if there is anything special at all in the
power-law distributions of node-activity, node-activity vectors, and
layer activity, and if the observed behaviour of multiplexity and
normalised Hamming distance among layers can be just the result of the
juxtaposition of independent layers. We propose here four different
multiplex network models and we compare the correlations in node and
layer activity observed in real-world multiplexes with those produced
by those models. The first three models are null-models to assess the
significance of the heterogeneity of the distributions
$P(N\lay{\alpha})$, $P(B_i)$ and $P(\bm{b}_i)$. 
The fourth model is instead a generative
model which proposes a possible explanation for the observed
distributions of pairwise multiplexity and normalised Hamming distance
among layers. A software implementation of the four models is
available for download at~\cite{url}.

\medskip
\textit{Hypergeometric model (HM). ---} In this model we fix the
numbers $N\lay{\alpha}$ of active nodes at each layer $\alpha$ to be
equal to those observed in the original multiplex network. The
$N\lay{\alpha}$ nodes to be actived at each layer $\alpha$ are then
randomly sampled with a uniform probability from the $N$ nodes of the
graph.  In this way, the activity of a node at a given layer is
uncorrelated from its activity at another layer and, given two layers
$\alpha$ and $\beta$, with $N\lay{\alpha}$ active at the first layer
and $N\lay{\beta}$ active at the second layer, the probability $p(m;
N, N\lay{\alpha}, N\lay{\beta})$ that exactly $m$ nodes, with
$m=0,\ldots, \min(N\lay{\alpha},N\lay{\beta})$, are active at both
layers follows a hypergeometric distribution:
\begin{equation}
  p\left(m; N, N\lay{\alpha}, N\lay{\beta}\right) =
  \frac{\displaystyle\binom{N\lay{\alpha}}{m} \binom{N-N\lay{\alpha}}{N\lay{\beta}
      - m}}{\displaystyle\binom{N}{N\lay{\beta}}}.
\end{equation}
Consequently, the average number of nodes active at both layers is
equal to ${N\lay{\alpha}N\lay{\beta}}/{N}$, and the expected pairwise
multiplexity of the two layers is:
\begin{equation}
  \widetilde{Q}_{\alpha,\beta} = \frac{N\lay{\alpha}N\lay{\beta}}{N^2}
\end{equation}
Similarly, the expected value of the normalised Hamming distance
between two layers $\alpha$ and $\beta$ is equal to:
\begin{equation}
  \widetilde{H}_{\alpha,\beta} = \frac{
    \displaystyle\sum_{m=0}^{N\lay{\beta}} \left(N\lay{\alpha} +
    N\lay{\beta} - 2m\right) \times p\left(m;N, N\lay{\alpha},
    N\lay{\beta}\right)} { \displaystyle \min(N, N\lay{\alpha} + N\lay{\beta})}
  \label{eq:hamming_expected}
\end{equation}

\begin{figure}[!t]
  \begin{center}
    \includegraphics[width=3in]{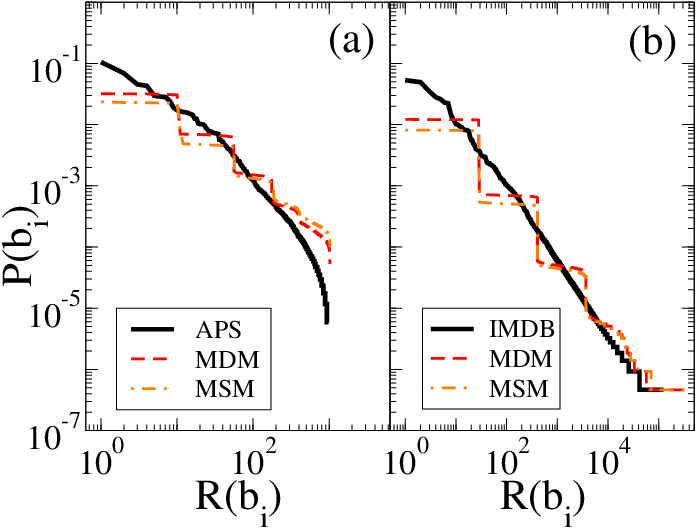}
  \end{center}
  \caption{(color online) The rank distribution of node-activity 
    vectors in APS (a) and IMDb (b), compared with those of synthetic
    multiplex networks genrated using MDM and MSM.}
  \label{fig:APS_IMDb_real_models}
\end{figure}

\textit{Multi-activity Deterministic Model (MDM). ---} In this model
we costruct networks with the same number of layers $M$ and the same
number of active nodes $N$ as in a given real-world multiplex network.
We consider a node active if it is active at at least one of the $M$
layers of the original network.  Then, we associate to each active
node $i$ a node-activity vector sampled at random among the
$\binom{M}{B_i}$ $M$-dimensional binary vectors having exactly $B_i$
non-zero entries, where $B_i$ is the number of layers in which node
$i$ is active in the original network. We name the model
Multi-activity Deterministic Model, since the distribution of $B_i$ of
the original multiplex is preserved, although the correlations in
layer activity and the distribution of node-activity vectors are
destroyed. The uniform assignment of node-activity vectors also
implies that all the layers will have, on average, the same number of
active nodes, since the probability that a given node $i$ is active on
a given layer $\alpha$ is equal to $B_i/M$ and does not depend on
$\alpha$. In particular, the expected number of active nodes at layer
$\alpha$ is:
\begin{equation}
  \widetilde{N}\lay{\alpha} = \frac{1}{M}{\sum_{i}B_i},\quad \forall
  \alpha.
\end{equation}

\medskip
\textit{Multi-activity Stochastic Model (MSM). ---} In this model, we
activate node $i$ at layer $\alpha$ with probability $\overline{B_i} =
B_i/M$, where $B_i$ is the node-activity of $i$ in the original network. 
Also in this case the expected activity of each layer is equal to
$M^{-1}\sum_i B_i$, but the node-activity of each node $i$ is a
binomially distributed random variable centered around $B_i$, so that, 
differently from MDM, the node-activity distribution is not preserved.

\medskip
\textit{Layer Growth with Preferential Activation Model (LGM). ---}
This model takes into account the fact that real-world multiplex
networks exhibit fat-tailed distributions of layer activity, and aims
at explaining the power-law distribution of node-activity reported in
Fig.~\ref{fig:multi_activity_real}. The main assumption of the model
which is certainly valid for some networks such as the continental
airlines, is that a multiplex network grows through the addition of
entire \textit{layers}, each arriving with a certain number of nodes
to be activated. Then, each node $i$ of a newly arrived layer is
activated (at that layer) with a probability that increases linearly
with the number of other layers in which $i$ is already active.  From
an operational point of view, we start from a multiplex consisting of
$N$ nodes (either active or inactive) and $M_0$ layers, and we add a
layer at each time step. Therefore, at time $t$ the multiplex has $M_0
+ t$ layers. We assume that in the newly arrived layer $\alpha$ there
are $N\lay{\alpha}$ nodes to be activated, where $N\lay{\alpha}$ is
set equal to the number of active nodes at layer $\alpha$ observed in
the original multiplex. Then, we consider all the nodes and activate
each node $i$ with probability:
\begin{equation}
  P_{i}(t) \propto A + B_i(t)
\end{equation}
where $A>0$ is a tunable real-valued parameter and $B_i(t)$ is the
number of layers (among the $M_0+t$ existing ones) in which
node $i$ is already active. The parameter $A$ is an intrinsic
attractiveness which guarantees that also nodes not yet
active in the existing layers have a non-zero 
probability of being activated at a new layer.   

\bigskip
In Figures ~\ref{fig:airports_real_models} 
and ~\ref{fig:APS_IMDb_real_models} we compare the results 
of the models with some measured quantities in real-world 
multiplex networks. In particular in Fig.~\ref{fig:airports_real_models}(a) 
we show the distribution of pairwise multiplexity 
for the European continental airlines and those 
obtained with the four synthetic models. Remarkably, the distribution
of multiplexity of the real system is pretty different from those
obtained through HM, MDM and MSM. In particular, both MDM and MSM
produce multiplex networks with an exponential-like distribution of
multiplexity, while in the original system $Q_{\alpha,\beta}$ is a
power-law. HM can somehow reproduce the
heterogeneity of $P(Q_{\alpha,\beta})$, even if the typical values of
$Q_{\alpha,\beta}$ are much smaller than those observed in the
European airline network. The best approximation is obtained through
the LGM, which reproduces quite accurately both the
shape and the slope of $P(Q_{\alpha,\beta})$.
Similarly, in Fig.~\ref{fig:airports_real_models}(b) we show the
distribution $P(B_i)$ of node-activity for the original European
airlines multiplex and the corresponding synthetic networks. Taking
aside MDM and MSM, for which the distribution of node-activity is
equal to that of the original network by construction, also in this
case LGM is the model which better approximates $P(B_i)$.
\begin{figure*}[!t]
  \begin{center}
    \includegraphics[width=6in]{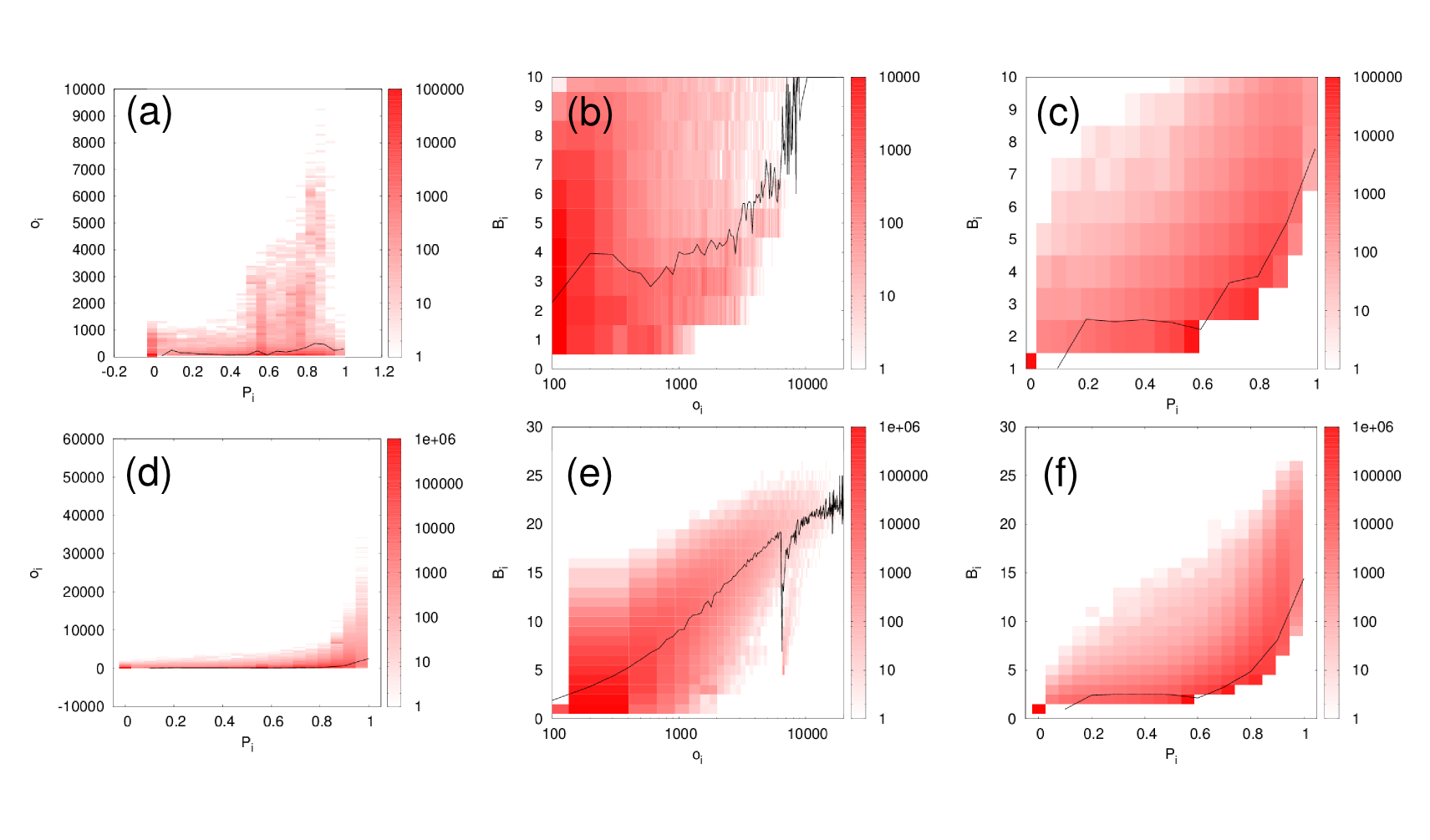}
  \end{center}
  \caption{(color online) Density plots of overlapping degree,
    participation coefficient and node-activity, for APS (top panels)
    and IMDb (bottom panels). On average, node-activity is positively
    correlated with both overlapping degree and participation
    coefficient (the solid line shows the average $\avg{B_i}$ computed
    over all the nodes having a certain value of $o_i$). However, the
    fluctuations in the values of $B_i$ are quite large in all the
    cases.}
  \label{fig:cartography}
\end{figure*}

Finally, in Fig.~\ref{fig:APS_IMDb_real_models} we compare the rank
distribution of node-activity vectors in APS and IMDb with those
obtained through MDM and MSM (we did not consider LGM since these
multiplex have a relatively small number of layers). We notice that
the Zipf's plots of the distributions produced by both MDM and MSM are
step-wise constant functions, in which each step corresponds to
node-activity vectors having the same value of non-null entries (i.e.,
of node-activity $B_i$). This is due to the fact that in MDM and MSM
the probability for a certain node-activity vector to be produced
depends only on the corresponding node-activity value $B_i$.

The results shown in Fig.~\ref{fig:APS_IMDb_real_models} suggest that
the pattern of node activity across layers in real-world multiplex
networks can be quite heterogeneous, and that indeed the activity of a
node at a certain layer is often highly correlated with its activity
(and inactivity) at other layers. This means that by studying the
properties of each layer separately, or, even worse, by aggregating
all layers in a single graph, one obtains only a partial picture of
the system, while a comprehensive understanding of a multi-layer
system requires to take into account the different layers altogether.

\section{Correlation between activity and degree}
\label{sec:cartography}

In this section we investigate the existence of correlations between
the activity of a node and its multidegree, i.e. the number of edges
incident in the node at each layer. To a first approximation, the
information contained in the multidegree of a node is well described
by only two quantities, the overlapping degree and the partecipation
coefficient of a node~\cite{Battiston2014}.  Following the definition
given in~\cite{Battiston2014}, we denote the \textit{overlapping
  degree} of node $i$ as:
\begin{equation}
  o_i=\sum_{\alpha} k\lay{\alpha}_i
\end{equation}
that is the total number of edges incident on $i$. Notice that $o_i$
is sometimes called the \textit{total degree} of node $i$. As the
degree is a proxy for the importance of a node in a single-layer
network, the overlapping degree of $i$ is a proxy for the overall
involvement of node $i$ in the multiplex network. However, the
overlapping degree measures only an aspect of the role played by a
node in a multiplex system. In fact, if we consider two nodes $i$ and
$j$, so that $i$ is active in all the $M$ layers and has $m$ links on
each of them, while $j$ is active only on one layer with $m\times M$
links, then we will have $o_i=o_j=m\times M$. Nevertheless, $i$ and
$j$ have quite different roles in the multiplex, since the removal of
node $j$ from the system will directly affect the structure of just
one layer (namely, the only layer in which $j$ is active), while the
removal of $i$ will potentially cause disruptions at all layers. In
order to quantify the heterogeneity of the distribution of the links
of a node across the layers, one can make use of the \textit{multiplex
  participation coefficient}~\cite{Battiston2014} :
\begin{equation}
P_i=\frac{M}{M-1}\left[1-
  \sum_{\alpha=1}^M\biggl(\frac{k_i^{[\alpha]}}{o_i}\biggr)^2\right].
\label{participationcoefficient}
\end{equation}
which takes values in $[0,1]$, is equal to $0$ if node $i$ is active
in exactly one layer, and tends to $1$ only if the edges of $i$ are
equally distributed across all the layers. It has been shown in
Refs.~\cite{Battiston2014,Nicosia2014} that important information on
the node properties of a multiplex can be obtained by a scatter plot
or a density plot of the participation coefficient as a function of
overlapping degree. Such diagram have been called \textit{multiplex
  cartography diagrams}. In Fig.~\ref{fig:cartography} panels (a) and
(d) we plot the multiplex cartography diagrams for APS and IMDb.
According to the values of the participation coefficient, nodes can be
divided into focused ($P_i<1/3$), mixed ($1/3<P_i<2/3$) and truly
multiplex ($2/3<P_i\le 1$). It is worth noticing that this
classification of nodes according to the value of their participation
coefficient is in line with the definition of network cartography
originally proposed in Ref.~\cite{Guimera2005} to characterise the
role played by single nodes in the organisation in communities, and
was adapted to multiplex networks in Ref.~\cite{Battiston2014}. More
principled ways to define the boundaries of the three regions might be
based, for instance, on the selection of percentiles of the
distribution of participation coefficients, e.g., by setting the
boundary between focused and mixed at the $50$-th percentile and the
boundary between mixed and multiplex at the $95$-th or $99$-th
percentile. However, such a choice would make difficult the comparison
of multiplex cartography diagrams associated to distinct multiplex
networks.

Nodes with relatively high values of $o_i$ are considered hubs.  By
construction, we do not expect a correlation between $o_i$ and $P_i$,
since the two quantities identify two different aspects of node
connectivity. And in fact, the diagrams shown in
Fig.~\ref{fig:cartography} exhibit a large variety of patterns. For
instance, APS is characterised by a relatively large fraction of mixed
hubs (nodes with high $o_i$ and intermediate values of $P_i$), while
almost all the hubs in the IMDb data set are truly multiplex (high
values of $P_i$).

In a similar way, we can quantify the existence of correlations
between the node-activity $B_i$ of a node $i$ and the corresponding
values of overlapping degree $o_i$ and participation coefficient
$P_i$.  In Fig.~\ref{fig:cartography} panel (b) and (e) we report the
density plots of node-activity and overlapping degree, respectively
for APS and IMDb. As expected, we observe positive correlations
between the two quantities $B_i$ and $o_i$, so that nodes with many
links tend to be active on more layers. This is reasonable because a
node with a small number of edges cannot be active on a large number
of layers.  However, the fluctuations around the average value of
node-activity for a certain value of overlapping degree (marked by the
black solid line in the plots) are quite large.
\begin{figure*}
  \begin{center}
    \includegraphics[width=6.2in]{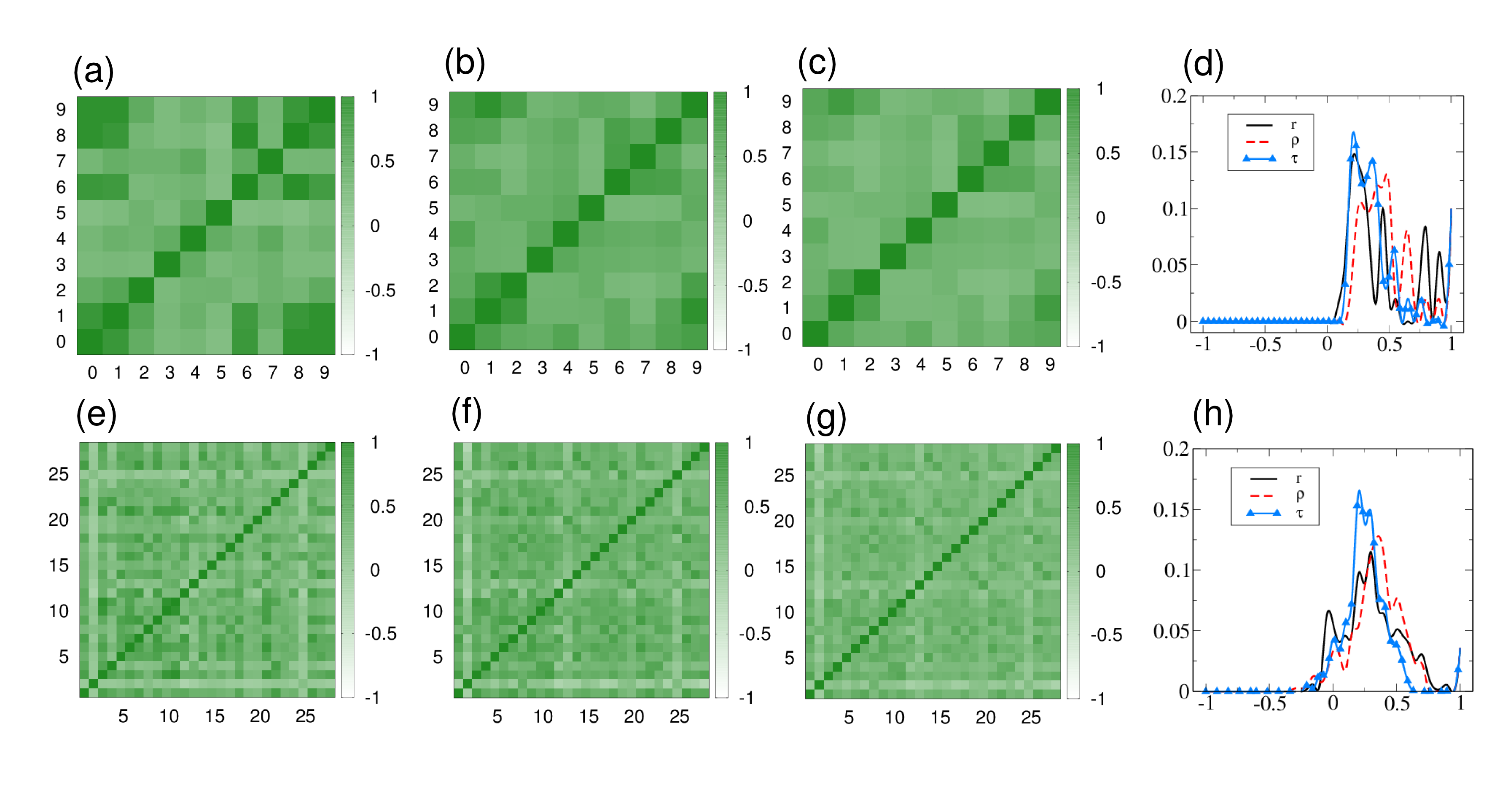}
  \end{center}
  \caption{(color online) Different degree correlation coefficients,
    namely (a) Pearson's $r$, (b) Spearman's $\rho$ and (c) Kendall's
    $\tau$ for different couples of layers, and the corresponding
    distributions (d) are reported for the APS, 
    and show that inter-layer
    correlations in this system tend to be assortative. A similar
    pattern is observed in IMDb (panels (e)-(h)). However, some movie
    genres, like Adult and Talk-Show (respectively corresponding to
    layer number 2 and number 25 in the diagram) have marked negative
    inter-layer correlations with almost all the other layers.}
  \label{fig:corr_coeffs}
\end{figure*}
Similar relationships exist between node-activity and participation
coefficient as shown in Fig.~\ref{fig:cartography}(c) and
Fig.~\ref{fig:cartography}(f), despite the existence of large
fluctuations. Namely, nodes having a higher value of participation
coefficient usually are active on more layers than nodes having small
values of $P_i$. This is indeed not surprising, sincethe edges of a
node with a higher value of participation coefficient are more
uniformly distributed across layers, hence the node will be active on
more layers.

\section{Inter-layer degree correlations}
\label{sec:degree-degree}

It has been extensively shown in the literature that single-layer
networks are characterised by the presence of degree-degree
correlations, meaning that nodes having a certain degree are
preferentially connected to other nodes having similar
(\textit{assortative} correlations) or dissimilar degree
(\textit{disassortative} correlations).  Social and communication
networks are the most remarkable examples of assortative networks,
while the vast majority of technological and biological networks
exhibit disassortative degree correlations. In addition to the
classical \textit{intra-layer} degree-degree correlations, in a
multiplex network we can also define the concept of
\textit{inter-layer} degree-degree correlations.

\medskip
\subsection{Inter-layer correlation coefficients}

A compact way to quantify the presence of inter-layer degree
correlations is to make use of one of the standard correlation
coefficients to measure how the degree sequences of two layers are
correlated. One possibility is the Pearson's linear correlation
coefficient~\cite{Parshani2010,Lee2012,Min2013_coop,Nicosia2014}. If
we denote as $k_i\lay{\alpha}$ and $k_i\lay{\beta}$ the degrees of
node $i$ respectively at layer $\alpha$ and layer $\beta$, the
Pearson's correlation coefficient of the two degree sequences is
defined as:
\begin{equation}
  r_{\alpha, \beta} = \frac{ \avg{ k_i\lay{\alpha} k_i\lay{\beta}} - 
\avg{k_i\lay{\alpha}} \avg{k_i\lay{\beta}}} {  
     \sigma_{k\lay{\alpha}}
     \sigma_{k\lay{\beta}}                   }
\end{equation}
To avoid the bias due to the relatively small multiplexity of
real-world systems, the averages are taken over all the nodes which
are active on both layers.
Another possibility is to use
the Spearman's rank correlation coefficient $\rho$~\cite{Nicosia2014}:
\begin{equation}
  \rho_{\alpha, \beta} = \frac{\sum_{i}\left(   R\lay{\alpha}_i -
        \overline{ R\lay{\alpha} }    \right) 
\left( R\lay{\beta}_i -
       \overline{ R\lay{\beta} }      \right)}
    {\sqrt{\sum_i\left(   R\lay{\alpha}_i -
        \overline{ R\lay{\alpha} }    \right)^2\sum_j \left(  
 R\lay{\beta}_j -  \overline{ R\lay{\beta} } 
\right)^2}}
\end{equation}
where $R\lay{\alpha}_i$ is the rank of node $i$ due to its degree on
layer $\alpha$, and $\overline{ R\lay{\alpha} }$ and $\overline{
  R\lay{\beta} }$ are the average ranks of nodes respectively at layer
$\alpha$ and layer $\beta$. Also in this case, only nodes active on
both layers are considered in the computation of
$\rho_{\alpha,\beta}$.
A third option is to use the Kendall's $\tau$ rank correlation
coefficient~\cite{Nicosia2014}:
\begin{equation}
  \tau_{\alpha,\beta}=\frac{n_c^{\alpha, \beta} -
    n_d^{\alpha,\beta}}{\sqrt{(n_0 - n_{\alpha})(n_0 - n_{\beta})}}.
\end{equation}
where $n_0=1/2 \times N Q_{\alpha, \beta} (NQ_{\alpha,\beta} - 1) $ ,
and $n_c^{\alpha, \beta}$ and $n_d^{\alpha, \beta}$ are, respectively,
the number of concordant pairs and the number of discordant pairs in
the two rankings. We say that the two nodes $i$ and $j$ are a
concordant pair if the ranks of the two nodes at the two layers agree,
i.e. if both $ R\lay{\alpha}_i > R\lay{\alpha}_j$ and $ R\lay{\beta}_i
> R\lay{\beta}_j$, or both $ R\lay{\alpha}_i < R\lay{\alpha}_j$ and $
R\lay{\beta}_i < R\lay{\beta}_j$.  If a pair of nodes is not
concordant, then it is said discordant.  Finally, $n_{\alpha}$ and
$n_{\beta}$ account for the number of rank ties in the two layers.

We have computed the three above pairwise correlation coefficients for
the APS and for the IMDb multiplex networks.  The results are shown in
Fig.~\ref{fig:corr_coeffs}. We notice that each of the three
coefficients show a slightly different behaviour. Nevertheless, it is
clear from the Figure that inter-layer correlations in APS are
exclusively assortative, while in IMDb we can observe both positive
and negative correlations. In particular, the degree of nodes at layer
2 (Adult movies) and at layer 25 (Talk-Shows) are negatively
correlated with the degree on all the other layers, whilst being
positively correlated to each other. These results indicate that it is
pretty uncommon ---even if not impossible--- for an actor of Adult
movies, to take part in a Family movie or in a Thriller. In addition
to this, the large majority of actors usually prefer to avoid
talk-shows, the main exception being porn stars. The presence of
negative inter-layer degree correlations in the IMDb multiplex network
is highlighted in the distributions of the three correlation
coefficients reported in panel (h). It is interesting that, in most of
the cases, also the inter-layer degree correlations in multiplex
social networks are assortative. This is in agreement with the common
belief that intra-layer degree-degree correlations in single-layer
social systems are always of the assortative type.  However, cases
such as the IMDb are an example that disassortativity is possible in
social networks when they are not aggregated, and treated as multiplex
networks.

It is important to stress that, although the Spearman's and Kendall's
rank correlation coefficients are able to capture, at least to some
extent, the presence of non-linear correlations in the rankings
induced by two degree distributions, the choice of which coefficient
is more appropriate to quantify inter-layer correlations might in
general depend on the actual system under study. As we will see in the
following, a more accurate way of measuring such correlations is by
means of inter-layer degree correlation functions.

\medskip
\subsection{Inter-layer correlation functions}

The complete information on degree correlations in single-layer
networks is contained in the joint degree distribution function
$P(k,k')$ or, equivalently, in the conditional degree distribution
$P(k'|k)$, which respectively denote the probability that a randomly
chosen link connects a node of degree $k$ to a node of degree $k'$,
and the probability that a link from a node of degree $k$ connects 
a node of degree $k'$. A convenient 
quantity that is commonly used to detect degree correlations is the
degree correlation function, defined as the average degree of the first
neighbours of a node having a certain degree $k$:
\begin{equation*}
  k_{nn}(k) =  \overline{k'}(k) = \sum_{k'}k' P(k'|k)
\end{equation*}
In fact, in single-layer networks with assortative degree correlations
$k_{nn}(k)$ will be an increasing function of $k$, while in
disassortative networks $k_{nn}(k)$ will decrease with $k$. An
interesting result is that in many cases of real-world complex
networks we have $k_{nn}(k) \sim k^{\nu}$, so that the correlation
exponent $\nu$ can be used to quantify the sign and intensity of
degree-degree correlations~\cite{Vespignani2001, Pastor-Satorras2001}.
\begin{figure}
  \begin{center}
    \includegraphics[width=3in]{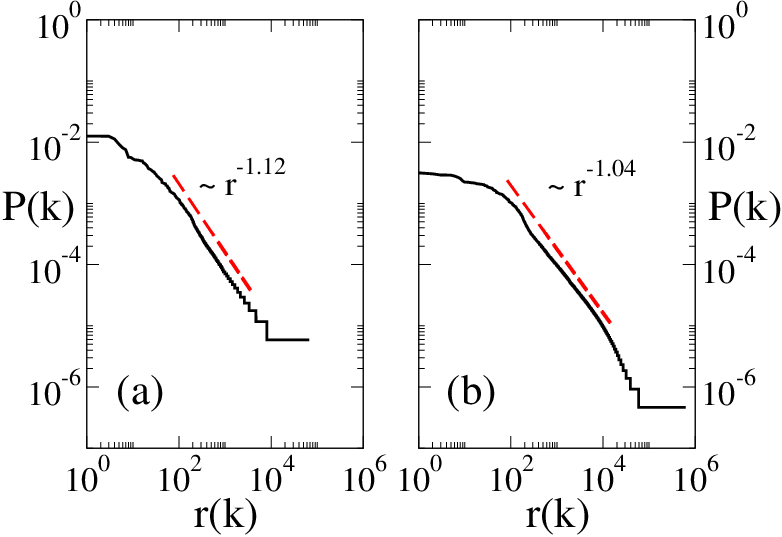}
  \end{center}
  \caption{(color online) The Zipf's plot of the distribution of
    multi-degree in (a) APS and (b) IMDb have a power-law tail with
    exponent close to $1.0$. However, the multi-degree distribution
    might be affected by large fluctuations. In fact, in both cases
    around $90\%$ of the multi-degree vectors are present only once,
    and more than $95\%$ are observed less than four times.}
  \label{fig:multideg_distr}
\end{figure}

\begin{figure*}[!t]
  \begin{center}
    \includegraphics[width=2in]{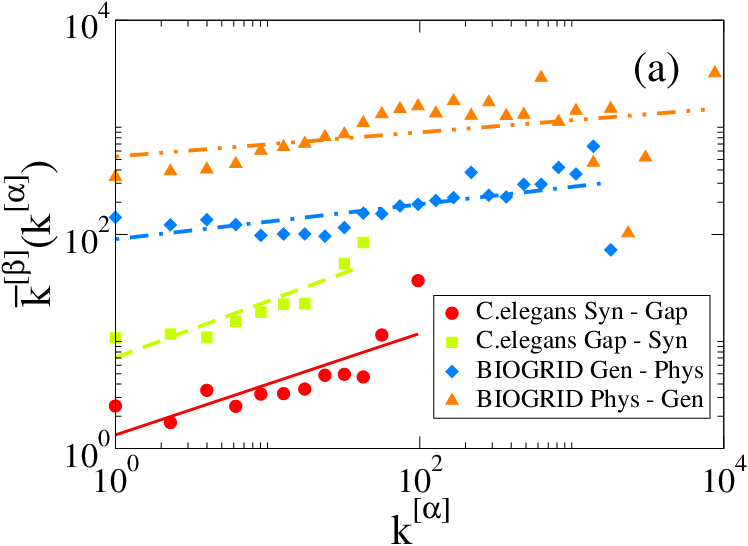}
    \includegraphics[width=2in]{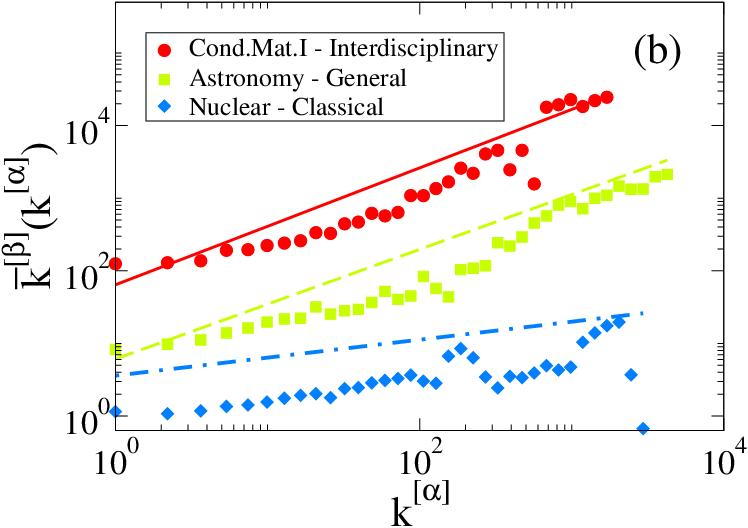}
    \includegraphics[width=2in]{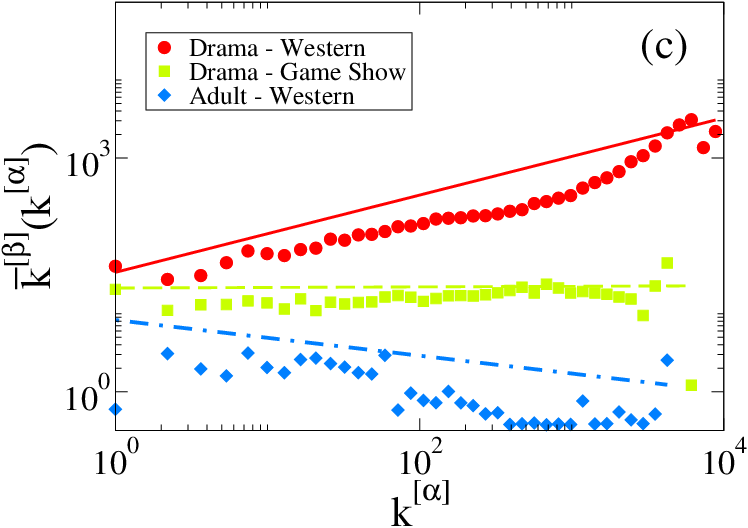}
  \end{center}
  \caption{(color online) The inter-layer pairwise degree correlation
    function $\overline{k\lay{\beta}}(k\lay{\alpha})$ is shown for (a)
    C.elegans and BIOGRID and for various couples of layers $\alpha$
    and $\beta$, respectively, in (b) APS and (c) IMDb. The lines
    reported are fit obtained by a power law of the form
    $\overline{k\lay{\beta}}(k\lay{\alpha})\sim
    ({k\lay{\alpha}})^{\mu}$. The plots are vertically displaced to
    enhance readability.
}
  \label{fig:corr_knn}
\end{figure*}

In a multiplex network the complete information about inter-layer
correlations is contained in the joint probability $P(k\lay{1},
\ldots, k\lay{M})$, which represents the probability that a randomly
chosen node has degree $k\lay{1}$ at layer 1, degree $k\lay{2}$ at
layer 2, and so on, and is nothing else than the multi-degree
distribution of the system $P(\bm{k})$.  As an example, we report in
Fig.\ref{fig:multideg_distr} the Zipf's plot of the distribution of
multi-degree for APS and IMDb. Interestingly, both distributions
exhibit a power-law behavior with a negative exponent around $-1.0$.
The inter-layer correlations between layers $\alpha$ and $\beta$ can
be studied by constructing the pairwise joint and conditional
probability distributions
\begin{equation*}
  P(k\lay{\alpha},k\lay{\beta}) \quad \text{and} \quad
  P(k\lay{\beta}|k\lay{\alpha})
\end{equation*}
The first quantity denotes the probability that a randomly chosen node
has degree $k\lay{\alpha}$ at layer $\alpha$ and degree $k\lay{\beta}$
at layer $\beta$, while the latter denotes the probability that a node
having a given degree $k\lay{\alpha}$ at layer $\alpha$ has degree
$k\lay{\beta}$ at layer $\beta$.  In the same spirit of the degree
correlation function defined for single-layer networks, given two
layers $\alpha$ and $\beta$ we can define the two inter-layer degree
correlation functions:
\begin{equation}
  \overline{k\lay{\beta}}(k\lay{\alpha}) = \sum_{k\lay{\beta}} k\lay{\beta}
  P(k\lay{\beta}|k\lay{\alpha})
\end{equation}
and
\begin{equation}
  \overline{k\lay{\alpha}}(k\lay{\beta}) = \sum_{k\lay{\alpha}}
  k\lay{\alpha} P(k\lay{\alpha}|k\lay{\beta}).
\end{equation}
These two quantities quantify the average degree at layer $\beta$
(resp. $\alpha$) of a node having a degree equal to $k\lay{\alpha}$
(resp. $k\lay{\beta}$) at layer $\alpha$ (resp. $\beta$). Being
average quantities, we expect smaller fluctuations than if we directly
plotted the two-dimensional functions $P(k\lay{\alpha},k\lay{\beta})$
and $P(k\lay{\beta}|k\lay{\alpha})$.  Again, the idea is that an
increase (decrease) of $\overline{k\lay{\beta}}(k\lay{\alpha})$ as a
function of $k\lay{\alpha}$ is a sign of the presence of assortative
(disassortative) inter-layer degree correlations between $\alpha$ and
$\beta$.

\begin{figure*}[!t]
  \begin{center}
    \includegraphics[width=1.8in]{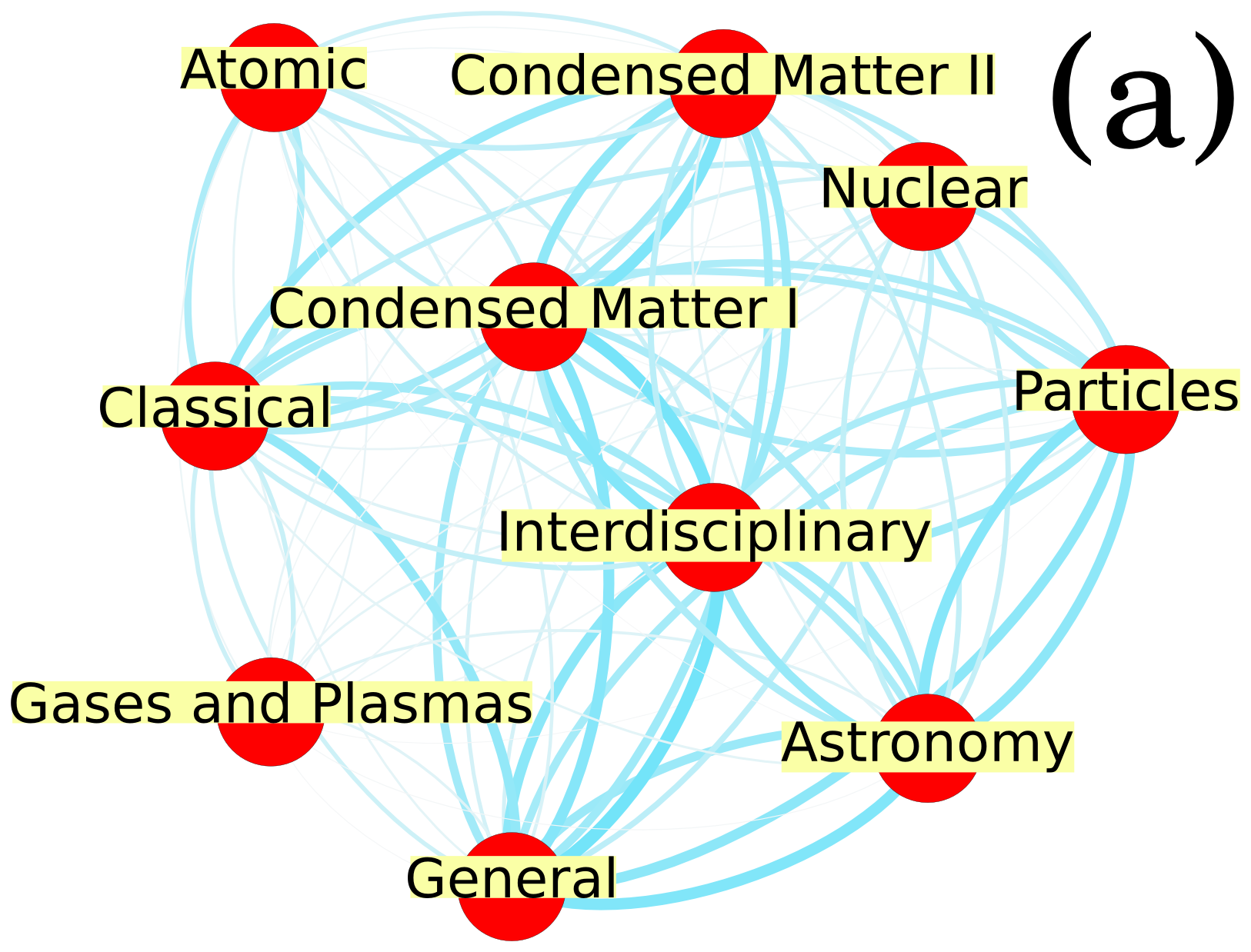}
    \includegraphics[width=2.1in]{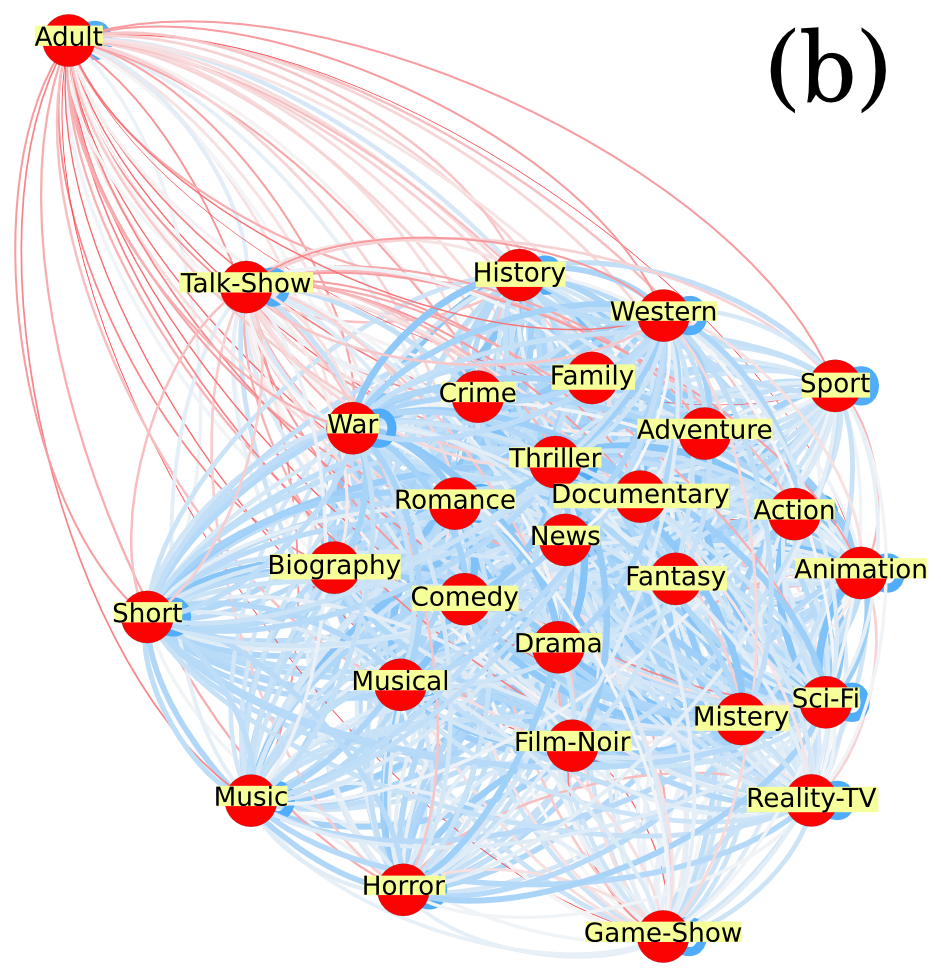}
    \includegraphics[width=2.2in]{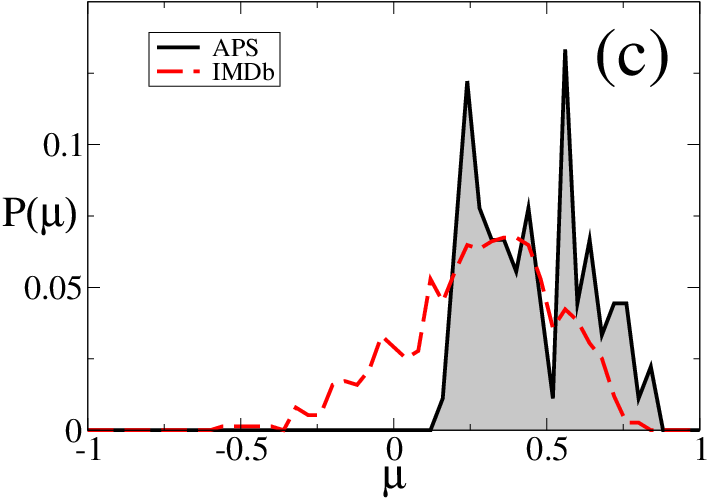}
  \end{center}
  \caption{(color online) The inter-layer correlation pattern of (a)
    APS and (b) IMDB is evident by considering a graph whose nodes
    correspond to layers and the weight of the edges is the value of
    the inter-layer correlation exponent $\mu$. In the figure blu
    weights correspond to positive correlations while red weights to
    negative ones. Panel (c): the distribution of the values of the
    inter-layer correlation exponent $\mu$ in APS (solid black line)
    and in IMBd (dashed red line). Notice that while inter-layer
    degree correlations are always positive in APS, the layers of IMDb
    might be either positively or negatively correlated.}
  \label{fig:corr_exp_distr}
\end{figure*}

In Fig.~\ref{fig:corr_knn} we show some typical examples of pairwise
inter-layer degree correlation functions in C.elegans, BIOGRID, APS
and IMDb. Both in the two biological networks and in APS we observe an
increasing behavior of $\overline{k\lay{\beta}}(k\lay{\alpha})$ as a
function of $k\lay{\alpha}$, denoting the presence of assortative
inter-layer degree correlations. For the multiplex network of movie
actor collaborations we find instead pairs of layers with assortative
or disassortative inter-layer degree correlations, and also pairs of
uncorrelated layers.  As an example of positively correlated genres in
the IMDb we report the couple Drama-Western.  The couple Adult-Western
is instead negatively correlated, while Drama Movies are not
correlated with Game Show, as witnessed by the fact that
$\overline{k\lay{\beta}}(k\lay{\alpha})$ shows no dependence on
$k\lay{\alpha}$.  It is worth noticing that also inter-layer
correlation functions can be well fitted, in most of the cases, by
power-laws in the form $\overline{k\lay{\beta}}(k\lay{\alpha})\sim
({k\lay{\alpha}})^{\mu}$, so that for each network, and for each
ordered pair of layers $(\alpha, \beta)$, it is possible to extract
the \textit{inter-layer correlation exponent} $\mu$.  We can therefore
say that we observe assortative, neutral or disassortative
correlations, depending on the fact that the sign of $\mu$ is
respectively positive, null or negative. The absolute value of $\mu$
then give information on the intensity of the correlations. Notice
that in general, according to the definition of
$\overline{k\lay{\alpha}}(k\lay{\beta})$, the exponent of
$\overline{k\lay{\beta}}(k\lay{\alpha})$ might be diffrent from the
esponent of $\overline{k\lay{\alpha}}(k\lay{\beta})$, as happens for
instance in Fig.~\ref{fig:corr_knn}(a) for the layers of C.elegans and
BIOGRID.

In Fig.~\ref{fig:corr_exp_distr} we report a graphical representation
of the inter-layer degree correlation patterns in APS and in IMDB and
we also show the corresponding distribution of inter-layer correlation
exponents observed in the two systems. Each node of the graphs shown
in Fig.~\ref{fig:corr_exp_distr}(a)-(b) corresponds to a layer of the
multiplex, and the color of a link represents the sign and magnitude
of the exponent of the inter-layer correlation function between two
layers (red for negative exponents and blue for positive ones).
It is evident that while in APS inter-layer degree correlations are
always positive, in IMDb they might be either positive or
negative. Notice also that the only layers in IMDB having negative
degree correlations with the others are those corresponding to Adult
movies and Talk-Shows.

\section{Models of inter-layer degree correlations}
\label{sec:model_degree-degree}

We propose here two different models to reproduce the observed
patterns of pairwise inter-layer degree correlations. The first model
is based on the tuning of the Spearman rank correlation coefficient
$\rho_{\alpha,\beta}$, while the second one allows to obtain an
inter-layer correlation function 
$\overline{k\lay{\beta}}(k\lay{\alpha})\sim
    ({k\lay{\alpha}})^{\mu}$.
with a prescribed value of the
correlation exponent $\mu$.  Both models are based on simulated annealing.

\subsection{Model for $\rho$}

Let us consider two graphs with the same number of nodes $N$.  If we
want to construct a two-layer multiplex network using the two graphs
respectively as layer $\alpha$ and layer $\beta$ of the multiplex, we
need to couple the nodes of the two graphs in such a way that each
node of layer $\alpha$ is connected with exactly one node on the other
layer $\beta$.  Such a coupling can be realized in many different
ways, and in particular it can be chosen in order to obtain a given
level of inter-layer degree correlation, for instance a given value of
the Spearman rank correlation coefficient $\rho_{\alpha,\beta}$. The
coupling/correspondence between the nodes of the two graphs can be
described by a $N \times N$ matrix $\mathcal{S}=\{s_{ij}\}$ that we
call \textit{assignment}. Entry $s_{ij}=1$ if node $i$ in layer
$\alpha$ corresponds to node $j$ in layer $\beta$. Since we have a
one-to-one correspondence between the nodes of the two graphs we have
to impose $\sum_{j}s_{ij}=1, \> \forall i$. For simplicity in the 
notation, let us denote by $x_i$ the rank of node $i$ in layer $\alpha$, as
induced by the degree sequence $\{k\lay{\alpha}_i\}$, and by $y_i$ the
rank of node $i$ in layer $\beta$, as induced by $\{k\lay{\beta}_i\}$.
In this case the Spearman's rank correlation coefficient corresponding
to the assignment $\mathcal{S}$ can be written as
\begin{equation}
  \rho = \frac{\sum_{i,j}s_{ij}(x_i -
    \bar{x})(y_j-\bar{y})}{\sqrt{\sum_{i}(x_i - \bar{x})^2
      \sum_{j}(y_j-\bar{y})^2}}
\end{equation}
This equation can be also expressed in the form
\begin{equation}
  \frac{\sum_{ij}s_{ij}x_iy_j + C}{D}
\end{equation}
where
\begin{equation}
  C = N\bar{x}\bar{y} - \bar{y}\sum_{i}x_i - \bar{x}\sum_{i}y_i
\end{equation}
and 
\begin{equation}
  D = \sqrt{\sum_{i}(x_i - \bar{x})^2 \sum_{j}(y_j-\bar{y})^2}
\end{equation}
are two constants which depends only on the two rankings $\{x_i\}$ and
$\{y_i\}$, and not on the actual assignment $\mathcal{S}$. Therefore, the
Spearman's correlation coefficient is uniquely determined by the term
$\sum_{i,j}s_{ij}x_iy_j$, i.e. by the adjacency matrix
$\mathcal{S}$. Consequently, one should in principle be able to obtain 
any prescribed value ${\rho}^*$ of the Spearman rank correlation
coefficient by appropriately changing the assignment, i.e. by finding
a matrix ${\mathcal{S}}^*=\{{s^*}_{ij}\}$ so that
\begin{equation}
  \frac{\sum_{i,j}{s^*}_{ij}x_i y_j + C}{D} = {\rho^*}
  \label{eq:eq_rho}
\end{equation}
For a generic assignment $\mathcal{S}$ we have:
\begin{equation*}
  \frac{\sum_{i,j}s_{ij}x_i y_j + C}{D} = \rho_{\mathcal{S}}
  \neq {\rho^*}
\end{equation*}
which is associated to the cost function
$F(\mathcal{S})=|\rho_{\mathcal{S}} - {\rho^*}|$. 

\begin{algorithm}[!t]
  \caption{Simulated annealing for ${\rho^*}$}
  \begin{algorithmic}[1]
    \REQUIRE{$\{\bm{k}_i\}$, $\mathcal{S} = \{s_{ij}\}$, ${\rho^*}$, $\varepsilon$} 
    \ENSURE{$\mathcal{S'} = \{s'_{ij}\}$ so that
      $\rho={\rho^*}$}
    \STATE{compute $\rho_{\mathcal{S}}$}
    \STATE{$F(\mathcal{S})$ $\gets$ $|\rho_{\mathcal{S}} - {\rho^*}|$}
    \WHILE{$ F(\mathcal{S}) > \varepsilon$}
    \STATE {select two inter-layer edges, $(i,j)$ and $(k,\ell)$, at random}
    \STATE {replace $(i,j)$ with  $(i,\ell)$ and $(k,\ell)$ with
      $(k,j)$}
    \STATE {compute $\rho_{\mathcal{S}'}$}
    \STATE {$F(\mathcal{S}')$ $\gets$ $|\rho_{\mathcal{S}'} - {\rho^*}|$}
    \IF {$F(\mathcal{S}') < F(\mathcal{S})$}
    \STATE {$\mathcal{S}$ $\gets$ $\mathcal{S}'$}
    \ELSE
    \STATE{swap $F(\mathcal{S})$ and $F(\mathcal{S}')$ with
      probability  $p=e^{-(F(\mathcal{S}') - F(\mathcal{S}))/\gamma}$}
    \ENDIF
    \STATE{$F(\mathcal{S})$ $\gets$ $|\rho_{\mathcal{S}} - {\rho^*}|$}
    \ENDWHILE
    \RETURN {$\mathcal{S}$}
  \end{algorithmic}
  \label{algo:tune_rho}
\end{algorithm}

The basic idea is then to subsequently modify the structure of the
assignment in order to minimize $F(\mathcal{S})$.  We will make use of
a simulated annealing algorithm which works as follows. We start from
an initial random assignment $\mathcal{S}$, and we compute its
associated cost function $F(\mathcal{S})$. Then, we select two edges
$e_1=(i,j)$ and $e_2 = (k,\ell)$ of $\mathcal{S}$, uniformly at random
so that $e_1\neq e_2$, and we consider the adjacency matrix associated
to the assignment $\mathcal{S}'$ obtained from $\mathcal{S}$ by
replacing $e_1$ and $e_2$ with $e_1'=(i,\ell)$ and $e_2' = (k,j)$. We
compute $F(\mathcal{S}')$, and we accept the new assignment
$\mathcal{S}'$ with a probability
\begin{equation}
  p = 
  \begin{cases}
    1 & \text{ if } F(\mathcal{S}') < F(\mathcal{S})
    \\ e^{-\frac{F(\mathcal{S}') - F(\mathcal{S})}{\gamma}} &
    \text{otherwise}
  \end{cases}
\end{equation}
where $\gamma$ is a parameter. This scheme, whose pseudo-code is
reported in Algorithm~\ref{algo:tune_rho}, will favour changes to the
adjancency matrix which contribute to minimize the function $F$, but
it also allows to explore ergodically all the possible configurations
of $\mathcal{S}$, by accepting unfavourable changes with a finite
probability.  Notice that, due to the discrete nature of the
assignment problem and depending on the characteristics of the two
rankings under consideration, it might happen that there exists no
assignment which produces exactly the desired value
${\rho^*}$. Consequently, the algorithm will stop when
$F(\mathcal{S})<\varepsilon$, where $\varepsilon$ is a threshold set
by the user. Moreover, in order to avoid any bias due to the
relatively small multiplexity of real-world systems (i.e., to the
relatively small fraction of nodes which are active on both $\alpha$
and $\beta$, for any choice of $\alpha$ and $\beta$), it is usually
better to run the algorithm only on the nodes which are active on both
the layers considered. In the generic case of $M$-layer multiplex
networks one can iterate this algorithm in order to set the values of
$\rho_{\alpha,\beta}$ for up to $M-1$ pairs of layers. 

\begin{figure*}
  \begin{center}
    \includegraphics[width=2in]{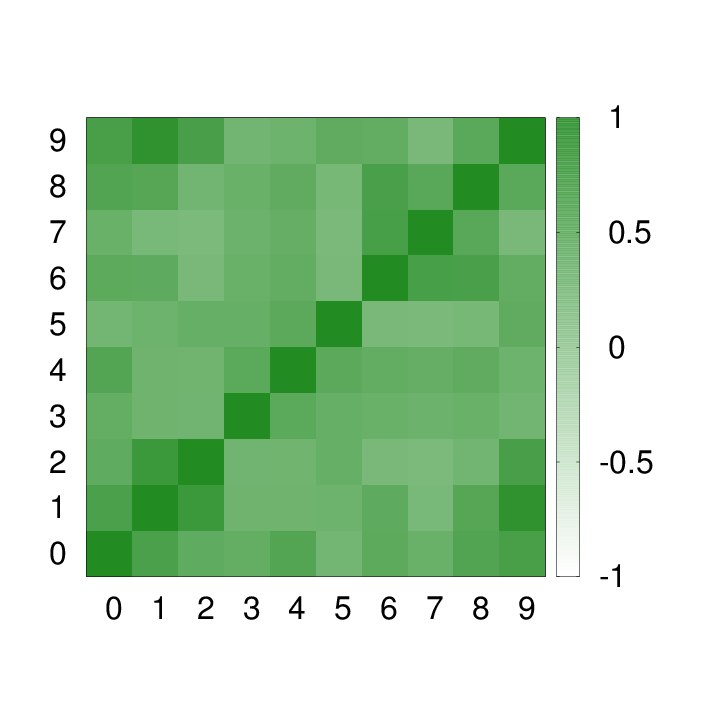}
    \includegraphics[width=2in]{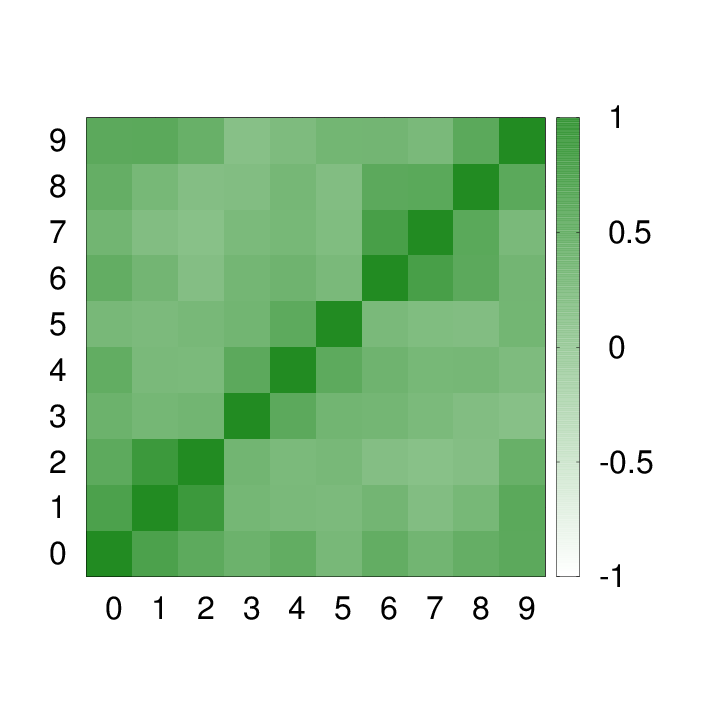}
    \includegraphics[width=2in]{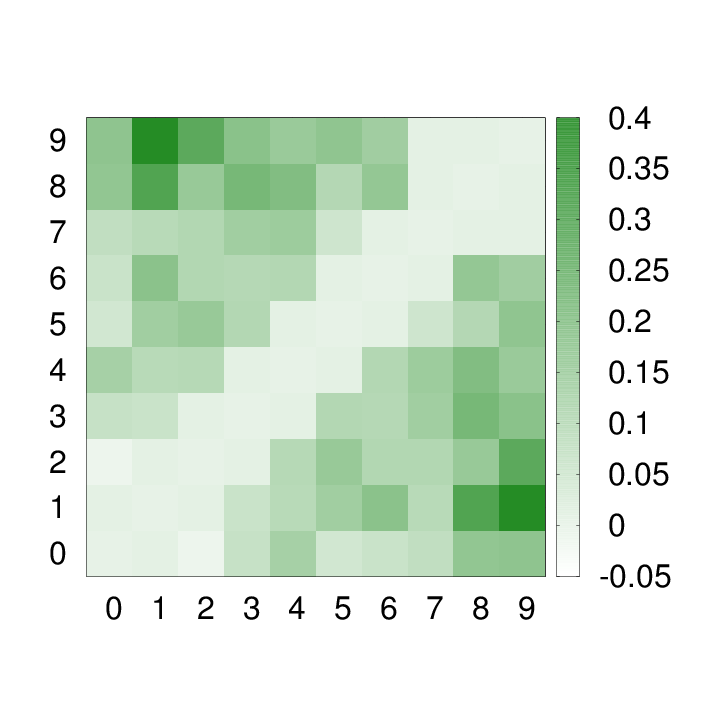}\\
    \includegraphics[width=2in]{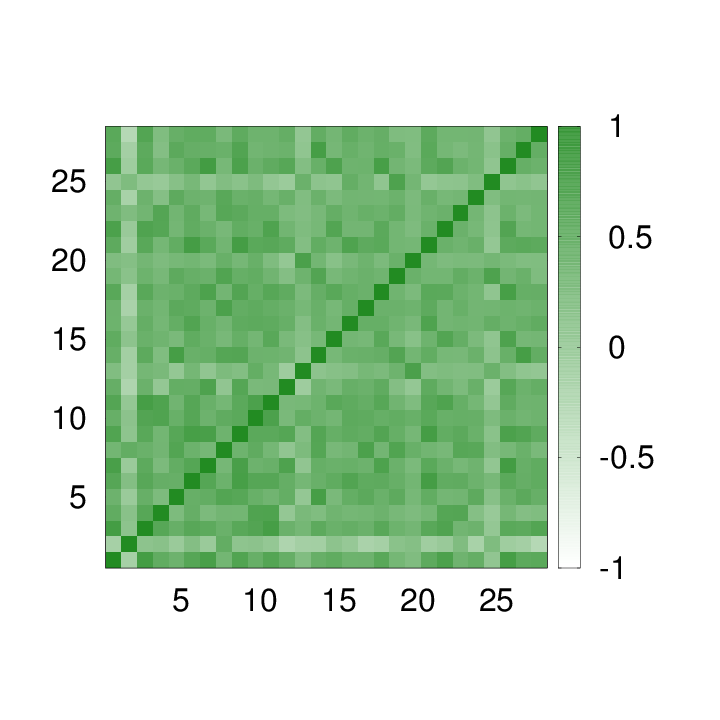}
    \includegraphics[width=2in]{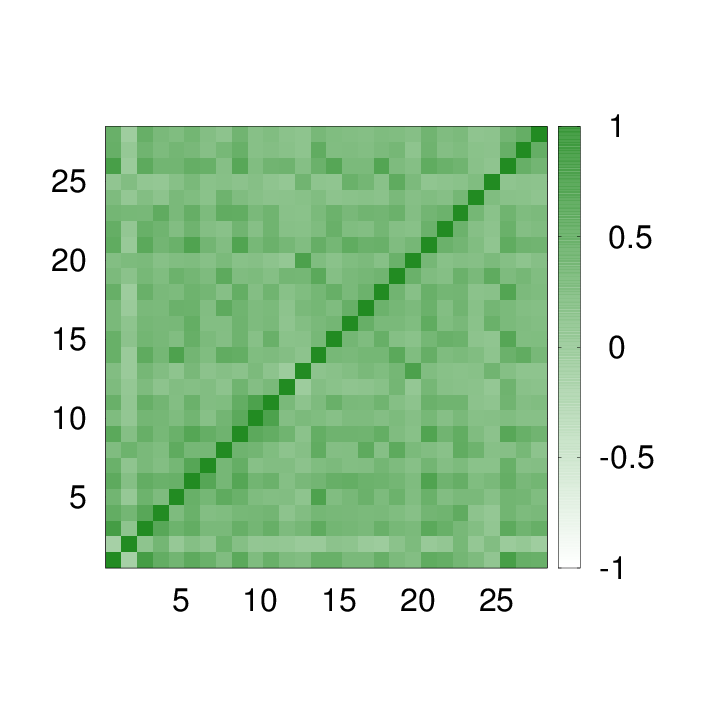}
    \includegraphics[width=2in]{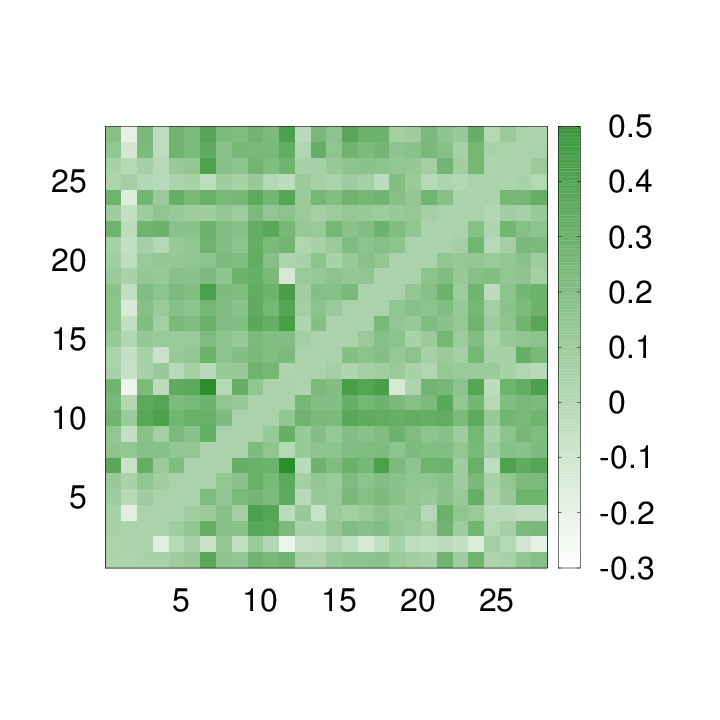}
  \end{center}
  \caption{(color online) The values of the Spearman correlation
    coefficient in the original multiplex (left panels) and in that
    obtained through Algorithm~\ref{algo:tune_rho} (middle panels)
    respectively for APS (top) and IMDb (bottom). In the rightmost
    panel we show the difference between the original distribution of
    $\rho$ and that obtained in the synthetic network. In both cases,
    the overall shape of the distribution of inter-layer correlations
    in the synthetic multiplex looks very similar to the original
    one. However, the differences in the obtained value of $\rho$
    might be quite high. This is due to the fact that
    Algorithm~\ref{algo:tune_rho} allows to set only $M-1$ pairs or
    correlations, over the total $M(M-1)/2$.}
  \label{fig:rho_model}
\end{figure*}

As an example, we report in Fig.~\ref{fig:rho_model} the values of
$\rho_{\alpha,\beta}$ measured for the APS and for the IMDb, together
with those obtained in the synthetic multiplex networks constructed by
using the proposed algorithm.  Each synthetic network was constructued
by keeping the distribution of node-activity vectors of the original
multiplex, and by reassigning at random the degrees of the active
nodes at each layer, sampling them from the same distribution observed
in the real multiplex.  We considered the $M-1$ pairs of layers having
consecutive IDs (e.g., couples of layers $(\alpha, \beta)$ such that
$\beta = \alpha+1$, for instance $(0,1)$, $(1,2)$ and so on), and we
measured the observed inter-layer rank correlation coefficients
$\rho_{\alpha,\beta}$.  Then, we iterated
Algorithm~\ref{algo:tune_rho}, starting from the first two layers,
setting ${\rho^*} = \rho_{\alpha,\beta}$ and obtaining an optimal
assignment of the nodes in $\alpha$ and $\beta$. Keeping fixed this
assignment, we run again Algorithm~\ref{algo:tune_rho} on the second
and the third layer of the multiplex, and we obtained the optimal
assignment between their nodes, and so forth. By looking at
Fig.~\ref{fig:rho_model} it is evident that there is a qualitative
correspondence between the distributions of $\rho$ in real and
synthetic networks, mostly due to the fact that partial ordering is a
transitive relation, but in general the difference between the two
might be relatively high (up to $0.4$ in APS and up to $0.5$ in IMDb).

It is important to stress here that Algorithm~\ref{algo:tune_rho} can
be straightforwardly generalized to work with any other node
property. In fact the algorithm is based on the comparison of rankings
induced by node properties, independently from the fact that these
rankings are induced by degree sequences or by any other node
attribute. Consequently, the same procedure can be employed to set the
magnitude and sign of inter-layer correlations with respect to any
real-valued pairs of node properties, such as the clustering
coefficient, the betweenness, or the size of the community to which a
node belongs. Notice that it would also be possible to consider
multi-objective functions which allow to set the correlations for all
the $M(M-1)/2$ pairs of layers at the same time. Such variants of
Algorithm~\ref{algo:tune_rho} will be the subject of another work
currently in preparation. A software implementation of
Algorithm~\ref{algo:tune_rho} is available for download at~\cite{url}.

\begin{algorithm}[!t]
  \caption{Simulated annealing for $\bar{q}=ak^{\mu}$}
  \begin{algorithmic}[1]
    \REQUIRE {$\{k_i\}$, $\{q_i\}$, $\{s_{ij}\}$, $\mu$, $\varepsilon$}
    \ENSURE {$\{s'_{ij}\}$ so that $|\bar{q} - ak^{\mu}|<
      \varepsilon$}
    \STATE {continue $\gets$ \textbf{True}}
    \WHILE {continue \textbf{is} \textbf{True}}
    \STATE {select two nodes, $i$ and $j$, at random}
    \STATE {$\Delta_{1}^{\rm old} \gets |\log(q_i) - \log(a) - \mu\log(k_i)|$}
    \STATE {$\Delta_{2}^{\rm old} \gets |\log(q_j) - \log(a) - \mu\log(k_j)|$}
    \STATE {$\Delta_{1}^{\rm new} \gets |\log(q_j) - \log(a) - \mu\log(k_i)|$}
    \STATE {$\Delta_{2}^{\rm new} \gets |\log(q_i) - \log(a) -
      \mu\log(k_j)|$}
    \STATE {$F_{\rm old} \gets \Delta_1^{\rm old} + \Delta_2^{\rm
        old}$}
    \STATE {$F_{\rm new} = \Delta_1^{\rm new} + \Delta_2^{\rm new}$}
    \IF{$F_{\rm new} < F_{\rm old}$}
    \STATE {swap $i$ and $j$ in the second layer and obtain $\{s'_{ij}\}$}
    \ELSE
    \STATE {swap $i$ and $j$ with  probability $p=e^{-(F_{\rm new} - F_{\rm old})/\gamma}$}
    \ENDIF
    \STATE {compute the best power-law fit $a'k^{\mu'}$ of $\bar{q}$}
    \IF {$|\mu - \mu'| < \varepsilon$ }
    \STATE {continue $\gets$ \textbf{False}}
    \ENDIF
    \ENDWHILE
    \RETURN {$\{s'_{ij}\}$}
  \end{algorithmic}
  \label{algo:tune_qnn}
\end{algorithm}

\subsection{Model for $\overline{k\lay{\beta}}(k\lay{\alpha})$}

Analogously to what done in the previous subsection, here we propose
an algorithm to tune the assignment of the nodes of two layers
$\alpha$ and $\beta$ in order to set a prescribed inter-layer degree
correlation function. In particular, we will assume that the desired
correlation function is a power-law,
i.e. $\overline{k\lay{\beta}}(k\lay{\alpha}) = a
({k\lay{\alpha}})^{\mu}$, as those observed in real-world multiplex
networks. To simplify the notation here we will indicate as $q$ the
degree of the node at layer $\beta$ and as $k$ the degree of the node
at layer $\alpha$. Then the desired correlation function has the form
$\overline{q}(k) = a k^{\mu}$ where the value of $\mu$ is that
obtained empirically for a given real network, while $a$ is a constant
to be determined.  The algorithm is similar to that proposed for the
adjustment of the Spearman's $\rho$ coefficient. We start from a
random assignment of nodes $\mathcal{S}$, we select two edges of
$\mathcal{S}$ uniformly at random and we try to swap their endpoints
in order to locally minimize the difference $\Delta$ between the
actual function $\overline{q}(k)$ and the desired one $k^{\mu}$.
Favorable swaps, i.e. those which produce smaller values of $\Delta$,
are always accepted, while unfavorable ones, i.e. those which produce
a local increase in $\Delta$, are accepted with a probability which
decays exponentially with the difference in $\Delta$.  The main steps
of the procedure are summarized in Algorithm~\ref{algo:tune_qnn}.
There are some technical subtleties to take into account for the
implementation of Algorithm~\ref{algo:tune_qnn}. First of all, the
fact that the coefficient $a$ which multiplies $k^{\mu}$ is in general
unknown. Consequently, $a$ is initially set to an arbitrary positive
value and then it is adaptively changed as the algorithm proceeds, by
setting it equal to the coefficient obtained through the best
power-law fit of $\bar{q}(k)$. Updates of $a$ are performed once every
$t_{a}$ steps of the algorithm, where $t_{a}$ is a parameter set by
the user. A software implementation of Algorithm~\ref{algo:tune_qnn}
is available at~\cite{url}.

\begin{figure*}
  \begin{center}
    \includegraphics[width=2in]{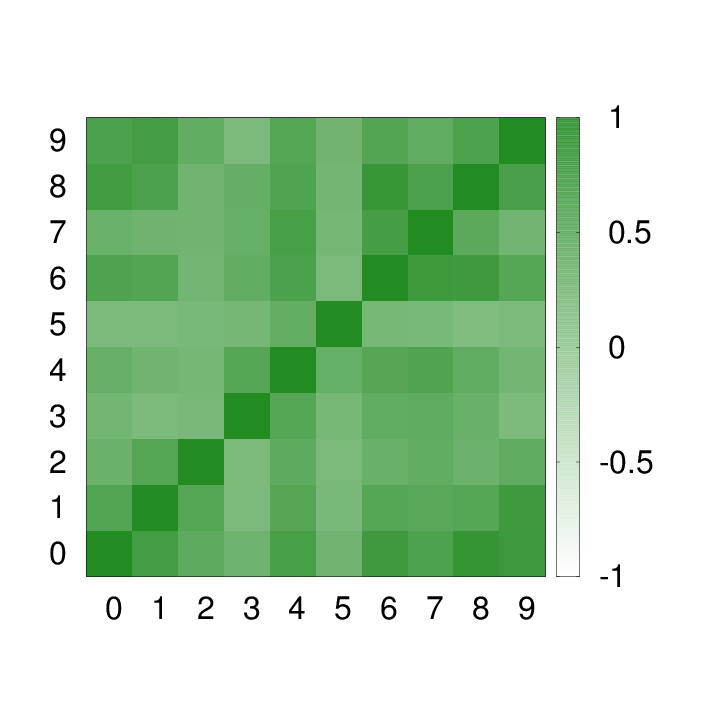}
    \includegraphics[width=2in]{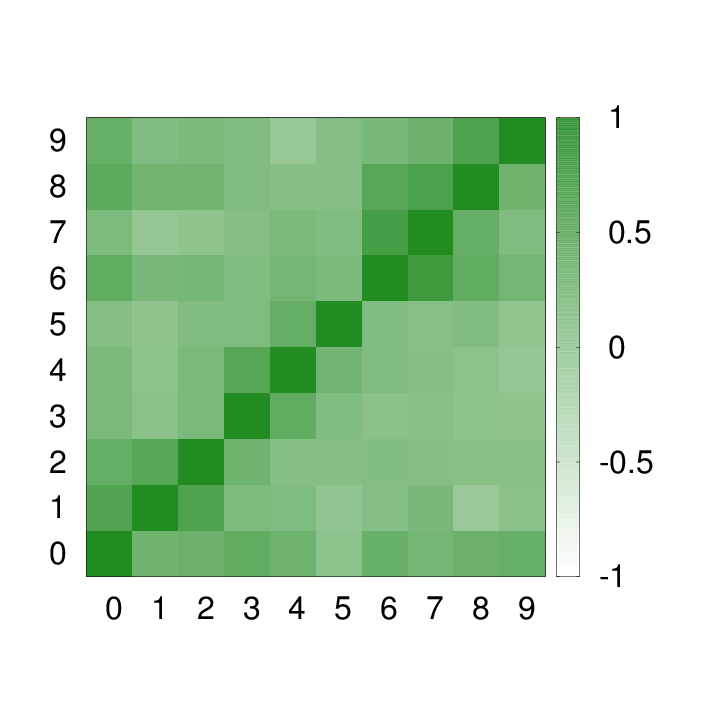}
    \includegraphics[width=2in]{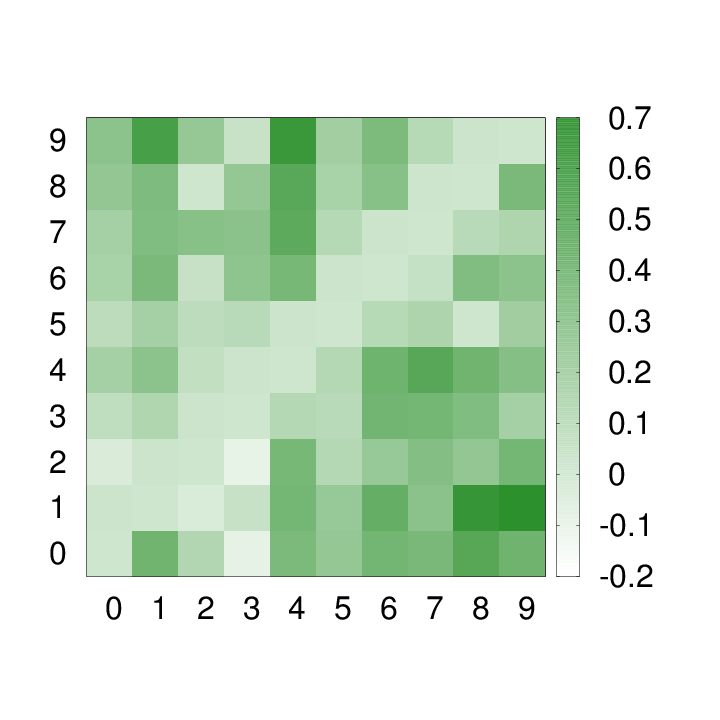}
  \end{center}
  \caption{(color online) The values of the inter-layer degree
    correlation exponent $\mu$ in the APS multiplex (left) and in a
    synthetic multiplex network generated through
    Algorithm~\ref{algo:tune_qnn} (middle). The rightmost panel shows
    the difference between the exponents observed in the original
    system and those measured in the synthetic network. Although the
    left and the middle panel look qualitatively similar, the right
    panel reveals that the difference in the actual inter-layer degree
    correlation exponent $\mu$ of the synthetic network might be as
    high as $0.7$.}
  \label{fig:tune_qnn}
\end{figure*}

In Fig.~\ref{fig:tune_qnn} we compare the values of the inter-layer
degree correlation exponent $\mu$ observed in the APS multiplex and in
the synthetic network obtained through
Algorithm~\ref{algo:tune_qnn}. Despite the distribution of $\mu$ in
the synthetic multiplex looks qualitatively similar to that of the
original system, the difference in the actual value of $\mu$ can be
quite large. Remember that by using Algorithm~\ref{algo:tune_qnn} one
can set the value of $\mu$ only for $M-1$ pairs of layers, so the poor
agreement of the pattern of correlation observed in the model with
that of the original system suggests that inter-layer degree
correlations of the APS multiplex network are not just due to the
superposition of pairwise inter-layer correlations.

\section{Conclusions}
\label{sec:conclusions}

In the last fifteen years complex networks theory has shed new lights
on the structure, organization, dynamics and evolution of complex
systems, providing a unifying framework to characterize and model
diverse natural and man-made systems. However, a complex network is
rarely an isolated object, since its constituent nodes can belong to
different systems at the same time and can be connected through a
variety of different relationships. Despite being still in its
infancy, the multiplex network approach, which consists in
representing the different kinds of relationships among nodes as
separate layers of a multi-layer graph, provides a promising framework
to understand and model the structure of multi-layer interconnected
systems.

In this work we have analyzed multiplex networks obtained from
real-world biological, technological and social systems, and spanning
a wide range of sizes. We showed that real-world multiplex networks
tend to be quite sparse, meaning that only a few nodes are active on
more than one layer, and that the patterns of presence and involvement
of nodes through the layers are characterized by inter-layer
correlations, as clearly shown by the heterogeneous distributions of
node-activity and by the non-trivial inter-layer degree correlation
functions.  The observation of such non-trivial patterns indicates
that a multiplex is more than the sum of its layers and cannot be
described by a single-layer network obtained by aggregating the
layers. Recent results in the field actually confirm that such
multiplex patterns play a fundamental role in many dynamical processes
taking place on multiplex networks, and can indeed be responsible of
completely new physical phenomena, unobservable in single-layer
projections~\cite{Radicchi2015,Diakonova2015}.
  
Finally, it is interesting to notice that the large majority of models
for multiplex networks proposed so far are based on the assumption
that each node of a multiplex is active at all layers, and that all
layers have the same number of nodes. In the light of the results of
this paper, these assumptions result too simplistic for the modeling
of real-world multiplex systems. Despite some recent attempts to take
into account heterogeneities of node and layer
activities~\cite{Cellai2015}, we believe that further research is
needed in this direction to better understanding the elementary
processes which might be responsible for the formation of such
interesting structural patterns.

\begin{acknowledgments}
  The authors acknowledge the support of the EU Commission through the
  project LASAGNE (FP7-ICT-318132).  This research utilized Queen
  Mary's MidPlus computational facilities, supported by QMUL
  Research-IT and funded by EPSRC Grant No.EP/K000128/1.
\end{acknowledgments}

\section*{APPENDIX}
\appendix

\renewcommand\theequation{{A-\arabic{equation}}}
\renewcommand\thetable{{A-\Roman{table}}}
\renewcommand\thefigure{{A-\arabic{figure}}}
\setcounter{figure}{0}

\section{Multiplex network data sets}

We provide here a detailed description of the data sets studied in the
paper, illustrating how the associated multiplex networks were
constructed. All the data sets are available for download
at~\cite{url}. Notice that, for most of the data sets considered, it
is also possible to associate a weight to each edge of the network,
measuring the strength of the corresponding interaction. However,
given that the focus of the current work is on the characterization of
correlations in the activity of nodes and in their degrees, we have
considered all these multiplexes as unweighted. A study of the
correlations between degrees and weights in multiplex networks can be
found for instance in Ref.~\cite{Menichetti2014}.

\textit{C.elegans. --- } The \textit{Caenorhabditis elegans} is a
small nematode, the first multicellular organism whose genome has been
completely sequenced~\cite{Celegans_genome}. Thanks to the fact that
its body is transparent, scientists have had the opportunity to study
with unprecedented accuracy each and every cell of the C.elegans, and
in particular its neural network, which is to date the only fully
mapped brain of a living organism~\cite{White1986}. The network,
consisting of 281 neurons and around two thousands connections among
them, was first analyzed as a complex network by Watts and Strogatz in
their seminal paper on small-world networks~\cite{Watts1998}, and has
since then been thoroughly
studied~\cite{Latora2001,Arenas2008,NicosiaPNAS2013}.  One important
aspect of this network, which has been not considered in most of the
analyses so far, is that the neurons can be connected either by a
chemical link, a synapse, or by an ionic channel, the so-called gap
junction. These two types of connection have completely different
dynamics and function. Consequently, the neural network of the
C.elegans can be naturally represented as a multiplex networks with
$N=281$ nodes and two layers, respectively for synapses and gap
junctions (see Fig.~\ref{fig:fig1}). Details of this multiplex, such as
the number $N\lay{\alpha}$ of active nodes at each layer, i.e. nodes
with at least one link at that layer, are shown in
Table~\ref{tab:celegans_biogrid}. In this particular network we have
two layers, hence $\alpha=1$ or $\alpha=2$.

\textit{Genetic-Protein Network. ---} As another example of biological
system we considered BIOGRID~\cite{BIOGRID}, a public database which
collects and makes available for research genetic and protein
interaction data from several organisms, including humans. The whole
data set consists of around 500000 registered interactions among
proteins in more than 40 different species. At the highest possible
level, such interactions may be of two distinct types, namely physical
and genetic. Two proteins A and B are said to interact physically if
they can establish a physical contact to form a larger complex C,
while they interact genetically if one of the two proteins, say A,
regulates B, i.e. if A can trigger the activation (or repression) of
the gene responsible for the production of B. It is worth mentioning
that a more fine-grained classification of gene-protein interactions
allows to identify up to seven different layers, as already shown in
Refs.~\cite{DeDomenico2015structural, DeDomenico2014muxviz}. 

Also in this case, the research has been mostly focused on the study
of the structural properties either of physical or of genetic
interactions among proteins. We propose to study here the protein
interaction networks as a multiplex network and, starting from the
BIOGRID data set, we have constructed a network with two undirected
and unweighted layers corresponding, respectively, to physical and
genetic interactions among proteins. The resulting multiplex networks
has $N=54549$ nodes, and the basic properties of the two layers are
summarized in Table~\ref{tab:celegans_biogrid}.

\textit{OpenFlight. ---} Another system which has been thoroughly
investigated as a single-layer complex network is the airport
transport system ~\cite{Barrat2004,Colizza2006}. In this case the
nodes of the network stand for airports and a link represents the
existence of at least one direct flight between two airports. More
fine-grained information about the airport transportation network has
been recently made
available~\cite{Cardillo2013_Airports,Cardillo2013_Emergence}.  Here we
use a data set of aerial routes provided by
OpenFlight~\cite{OpenFlight}, a collaborative free online tool which
allows to map flights all around the world. Registered users of the
website can upload information about their trips and share this
information with friends. The maintainers of the website made
available a dump of the data set which contains information about
59036 routes between 3209 airports operated by 531 different airlines
spanning the whole globe. For each route we have information about the
start point, the end point and the company which operates the
flight. Starting from this data set, we constructed $6$ different
multiplex networks. Each multiplex network represents the routes of a
continent (Africa, Asia, Europe, North America, Oceania, South
America) and consists of as many layers as airlines operating in that
continent. The active nodes on each layer are the airports from which
the corresponding airline company has at least one flight, and links
represent the routes provided by that airline. In
Table~\ref{tab:OpenFlights} we report the basic features of each of
the six continental multiplexes.
\begin{table}
  \begin{tabular}{|c|c|c|c|c|}
    \hline
    \textbf{Network} & \textbf{M} & $\bm{\eta}$ & p-value \\
    \hline
    Africa & 84 & $1.64\pm 0.16$ & 0.20 \\
    Asia & 213 & $1.71\pm 0.12$  & 1.00 \\
    Europe & 175 & $1.48\pm 0.11$  & 0.11 \\
    North America & 143 & $1.52 \pm 0.12$ & 0.99  \\
    Oceania & 37 & $1.37 \pm 0.10$ & 0.04 \\
    South America & 58 & $1.58\pm 0.18$ & 0.90  \\
    \hline
  \end{tabular}
  \caption{For each of the six continental airplane multiplex networks
    constructed from the OpenFlight database, we report the number of
    layers $M$ and the exponent $\eta$ of the distribution
    $P(N\lay{\alpha})\sim (N\lay{\alpha})^{-\eta}$ of the number of
    non-isolated nodes in each layer. The values of $\eta$ and the
    corresponding standard deviations were determined using the
    maximum likelihood estimator for power-law
    distributions~\cite{Clauset2007}, while the p-value is based on
    the maximization of the Kolmogorov-Smirnov statistics over 1000
    bootstrapped realizations (higher values of p-value are more
    significant).}
  \label{tab:OpenFlights}
\end{table}

\textit{APS Coauthorship. ---} Coauthorship networks are commonly
constructed by connecting with an edge two researchers if they have
published one or more papers together.  We used a data set made
available by the American Physical Society (APS) which reports
information about all the papers published in any of the journals
edited by APS since 1893 and up to 2009. In this data set, each paper
published after 1975 is associated to up to four numeric codes, in the
format XX.YY.ZZ, which identify a sub-field or research area according
to the Physics and Astronomy Classification Scheme (PACS). At the
highest level, PACS codes are organized into ten groups, respectively
corresponding to sub-fields of physics. Starting from this data set,
we constructed a multiplex collaboration network consisting of 10
layers, in which nodes represent authors and links connect authors
having co-authored at least one paper. Authors with identical first
and last name are considered the same authors. Please refer to
Ref.~\cite{disambiguation} for a more comprehensive introduction on
the problem of disambiguating authors in collaboration networks. Each
layer corresponds to the collaborations identified by papers whose
PACS codes are in one of the 10 high-level categories. In
Table~\ref{tab:APS} we report the properties of the layers of this
network. Each layer has up to around $79000$ active nodes, and the
density varies across layers, according to the typical publication
policy of each area of physics. For instance, papers in condensed
matter and interdisciplinary physics are usually authored by just a
few authors, while papers produced by large collaborations, including
up to several hundred authors, are typical in particle physics,
nuclear physics and astronomy.

\begin{table}[!ht]
  \begin{tabular}{|c|l|r|r|r|}
    \hline
    \textbf{Layer} & \textbf{Field} & $N\lay{\alpha}$ &
    $K\lay{\alpha}$ & $\avg{k\lay{\alpha}}$\\
    \hline
    0 & General & 53170 & 1268045 & 47.7\\
    1 & Particles & 37861 & 4865557 & 257.0 \\
    2 & Nuclear & 32792 & 1747892 & 106.6\\
    3 & Atomic & 33649 & 189674 & 11.27\\
    4 & Classical & 40269 & 222328 & 11.04 \\
    5 & Gases and Plasmas & 14237 & 179786 & 25.3 \\
    6 & Condensed Matter I & 63560 & 611765 & 19.3\\
    7 & Condensed Matter II & 79416 & 631159 & 15.9  \\
    8 & Interdisciplinary & 45385 & 509058 & 22.4\\
    9 & Astronomy & 31540 & 2467703 & 156.5\\\hline\hline
    - &  Aggregated & 170397 & 6950611 & 81.6\\ 
    \hline
  \end{tabular}
  \caption{The APS multiplex collaboration network
    consists of ten layers, one for each field of physics. For each
    layer $\alpha$ we report the number of active nodes
    $N\lay{\alpha}$, the number of edges $K\lay{\alpha}$ and 
    the average degree $\avg{k\lay{\alpha}}$. We also report for reference 
    the number of nodes, the number of edges and the average degree of the
    single-layer network obtained by aggregating all the layers.}
  \label{tab:APS}
\end{table}

\textit{IMDb. ---} The Internet Movie Database (IMDb)~\cite{IMDb} is a
Web site providing comprehensive information about all the movie
productions around the world. The data set is maintained and updated
by volunteers, and made available for research use. It contains
information about casts, producers, directors, etc.~of several million
movies belonging to 30 different genres.  We constructed a multiplex
network of collaborations among actors in which nodes represent actors
and an edge exists between two nodes if the corresponding actors have
co-acted in at least one movie. Each of the 30 categories represents a
layer of the multiplex, so that if two actors have played a role in
the same horror movie, they will be connected by an edge at the
corresponding layer. In Table~\ref{tab:IMDb} we show the basic
characteristics of each layer of the multiplex. Notice that only 28 of
the 30 layers are reported, since two of the layers, namely those
corresponding to Experimental and Lifestyle movies, were deliberately
left out of this study, since they contained less than 20 actors
each. Notice also the wide variety of ranges in the number of active
nodes. For instance Film-Noir has about 7 thousand active nodes, while
Drama, have more than one million active nodes and more than 43
millions edges.
\begin{table}[!ht]
  \begin{tabular}{|c|l|r|r|r|}
    \hline
    \textbf{Layer} & \textbf{Genre} &  $N\lay{\alpha}$ &
    $K\lay{\alpha}$ & $\avg{k\lay{\alpha}}$\\ \hline
    1 & Action & 330333 & 11800436 &  71.4\\
    2 & Adult & 66756 & 1691208 & 50.7\\
    3 & Adventure & 210293 & 7390148 & 70.3 \\
    4 & Animation & 55376 & 1120523 & 40.5\\
    5 & Biography & 128552 & 4272197 & 66.5\\
    6 & Comedy & 810693 & 30118775 & 74.3 \\
    7 & Crime & 297554 & 10051325 & 67.6\\
    8 & Documentary & 313019 & 6850670 & 43.8 \\
    9 & Drama & 1091789 & 43352371 & 79.4\\
    10 & Family & 198301 & 5432262 & 54.8\\
    11 & Fantasy & 176080 & 5096872 & 57.9\\
    12 & Film-Noir & 7035 & 399548 & 113.6\\
    13 & Game-Show & 15222 & 282942 & 37.2\\
    14 & History & 124803 & 4137162 & 66.3\\
    15 & Horror & 263290 & 5428250 & 41.2\\
    16 & Musical & 121471 & 4118346 & 67.8\\
    17 & Music & 165110 & 4977063 & 60.3\\
    18 & Mystery & 168898 & 4226618 & 50.0\\
    19 & News & 21530 & 406166 & 37.7\\
    20 & Reality-TV & 29112 & 465244 & 32.0\\
    21 & Romance & 364042 & 13325687 & 73.2\\
    22 & Sci-Fi & 164468 & 4147689 & 50.4\\
    23 & Short & 644430 & 5117780 & 15.9\\
    24 & Sport & 101006 & 3643330 & 72.1\\
    25 & Talk-Show & 19700 & 516943 & 52.5\\
    26 & Thriller & 356776 & 10757551 & 60.3\\
    27 & War & 118960 & 3967033 & 66.7\\
    28 & Western & 56638 & 2101057 & 74.2\\\hline\hline
    - & Aggregated & 2158300 & $\sim 1.2\times 10^8$ & $\sim 100$ \\ 
    \hline
  \end{tabular}
  \caption{Basic features on each of 28 layers of the IMDb multiplex
    network and of the corresponding aggregated network. In this case,
    each layer corresponds to a movie genre.}
  \label{tab:IMDb}
\end{table}

\end{document}